\documentclass[structabstract]{aa}  
\usepackage{graphicx,natbib}
\usepackage{lscape}
\usepackage{longtable}
\usepackage[varg]{txfonts}

\newcommand{\mi}{\relax \ifmmode {\mu{\mbox m}}\else $\mu$m\fi}
\newcommand{\hii}{\relax \ifmmode {\mbox H\,{\scshape ii}}\else H\,{\scshape ii}\fi}
\newcommand{\sii}{\relax \ifmmode {\mbox S\,{\scshape ii}}\else S\,{\scshape ii}\fi}
\newcommand{\nii}{\relax \ifmmode {\mbox N\,{\scshape ii}}\else N\,{\scshape ii}\fi}
\newcommand{\neii}{\relax \ifmmode {\mbox Ne\,{\scshape ii}}\else Ne\,{\scshape ii}\fi}
\newcommand{\neiii}{\relax \ifmmode {\mbox Ne\,{\scshape iii}}\else Ne\,{\scshape iii}\fi}
\newcommand{\oiii}{\relax \ifmmode {\mbox O\,{\scshape iii}}\else O\,{\scshape iii}\fi}
\newcommand{\oii}{\relax \ifmmode {\mbox O\,{\scshape ii}}\else O\,{\scshape ii}\fi}
\newcommand{\oi}{\relax \ifmmode {\mbox O\,{\scshape i}}\else O\,{\scshape i}\fi}
\newcommand{\ha}{\relax \ifmmode {\mbox H}\alpha\else H$\alpha$\fi}

\newcommand{\hep}{\relax \ifmmode {\mbox H}\epsilon\else H$\epsilon$\fi}
\newcommand{\hdel}{\relax \ifmmode {\mbox H}\delta\else H$\delta$\fi}
\newcommand{\hgam}{\relax \ifmmode {\mbox H}\gamma\else H$\gamma$\fi}

\newcommand{\pa}{\relax \ifmmode {\mbox Pa}\alpha\else Pa$\alpha$\fi}
\newcommand{\hb}{\relax \ifmmode {\mbox H}\beta\else H$\beta$\fi}
\newcommand{\rdostres}{\relax \ifmmode {\,\mbox{R}}_{\rm 23}\else \,\mbox{R}$_{\rm 23}$\fi}
\newcommand{\ergs}{\relax \ifmmode {\,\mbox{erg\,s}}^{-1}\else \,\mbox{erg\,s}$^{-1}$\fi}
\newcommand{\me}{\relax \ifmmode {\,}^{-1}\else \,$^{-1}$\fi}

\newcommand{\msun}{\relax \ifmmode {\,\mbox{M}}_{\odot}\else \,\mbox{M}$_{\odot}$\fi}

\newcommand{\cmtres}{\relax \ifmmode {\,\mbox{cm}}^{-3}\else \,\mbox{cm}$^{-3}$\fi}
\newcommand{\cmdos}{\relax \ifmmode {\,\mbox{cm}}^{-2}\else \,\mbox{cm}$^{-2}$\fi}
\newcommand{\cmseis}{\relax \ifmmode {\,\mbox{cm}}^{-6}\else \,\mbox{cm}$^{-6}$\fi}
\newcommand{\hi}{\relax \ifmmode {\mbox H\,{\scshape i}}\else H\,{\scshape i}\fi}

\newcommand{\Spi}{{\it Spitzer}}
\usepackage[colorlinks]{hyperref}
\begin{document}

   \title{Spectral Energy Distributions of H\,{\sc ii} Regions in M\,33 ({{\tt HerM33es}}) }
   \author{
     M.\,Rela\~{n}o\inst{1}  \and
     S.\,Verley\inst{1} \and
     I. P\'erez\inst{1} \and
     C.\,Kramer\inst{2} \and
     D.\,Calzetti\inst{3} \and
     E.\,M.\,Xilouris\inst{4} \and
     M.\,Boquien\inst{5} \and
     J.\,Abreu-Vicente\inst{2} \and
     F.\,Combes\inst{6} \and
     F.\,Israel\inst{7} \and
     F.\,S.\,Tabatabaei\inst{8}\and
     J.\,Braine\inst{9} \and
     C.\,Buchbender\inst{2} \and           
     M. Gonz\'alez\inst{2} \and
     P.\,Gratier\inst{8} \and     
      S.\,Lord\inst{10} \and
      B. Mookerjea\inst{11} \and
      G.\,Quintana-Lacaci\inst{12} \and 
        P. van der Werf\inst{7}         }
   \institute{
     Dept. F\'{i}sica Te\'orica y del Cosmos, Universidad de Granada, Spain -- \email{mrelano@ugr.es}
     \and
     Instituto Radioastronom\'{i}a Milim\'etrica (IRAM), Av. Divina Pastora 7, N\'ucleo Central, E-18012 Granada, Spain
     \and
     Department of Astronomy, University of Massachusetts, Amherst, 
     MA 01003, USA 
     \and
     Institute for Astronomy, Astrophysics, Space Applications \& Remote Sensing, National Observatory of Athens, P. Penteli, 15236, Athens, Greece
     \and
     Laboratoire d'Astrophysique de Marseille, UMR 6110CNRS, 13388, Marseille, France
     \and
      Observatoire de Paris, LERMA, 61 Av. de l'Observatoire, 
     75014 Paris, France
     \and
     Leiden Observatory, Leiden University, PO Box 9513, NL 2300 RA Leiden, The Netherlands
      \and
     Institut de Radioastronomie Millim\'{e}trique, 300 rue de la Piscine, 38406 Saint Martin dÕHe\'{e}res, France
     \and
     Laboratoire d'Astrophysique de Bordeaux, Universit\'{e} Bordeaux 1, 
     Observatoire de Bordeaux, OASU, UMR 5804, CNRS/INSU, B.P. 89, 
     Floirac F-33270, France
     \and
     IPAC, MS 100-22 California Institute of Technology, Pasadena, CA 91125, USA
     \and
     Tata Institute of Fundamental Research, Homi Bhabha Road, Mumbai, 4000005, India 
      \and
     CAB, INTA-CSIC, Ctra de Torrej\'on a Ajalvir, km 4, EÐ28850 Torrej\'on de Ardoz, Madrid, Spain
                }

   \date{Received ; accepted }
  \abstract
   {}
   {Within the framework of the {{\it Herschel}} M\,33 extended survey {{\tt HerM33es}} and in combination with multi-wavelength data we study the Spectral Energy Distribution (SED) of a set of \hii\ regions in the Local Group Galaxy M\,33 as a function of the morphology. We analyse the emission distribution in regions with different morphologies and present models to infer the \ha\ Emission Measure observed for \hii\ regions with well defined morphology.}
   {We present a catalogue of 119 \hii\ regions morphologically classified: 9 filled, 47 mixed,  36 shell, and 27 clear shell \hii\ regions. For each object we extract the photometry at twelve available wavelength bands, covering a wide wavelength range from FUV-1516\,\AA\ (GALEX) to IR-250\,\mi\ ({{\it Herschel}}) and we obtain the SED for each object. We also obtain emission line profiles in vertical and horizontal directions across the regions to study the location of the stellar, ionised gas, and dust components. We construct a simple geometrical model for the clear shell regions, whose properties allow us to infer the electron density of these regions.}
   {We find trends for the SEDs related to the morphology of the regions, showing that the star and gas-dust configuration affects the ratios of the emission in different bands. The mixed and filled regions show higher emission at 24\,\mi, corresponding to warm dust, than the shells and clear shells. This could be due to the proximity of the dust to the stellar clusters in the case of filled and mixed regions. The far-IR peak for shells and clear shells seems to be located towards longer wavelengths, indicating that the dust is colder for this type of objects.The logarithmic 100\,\mi/70\,\mi\ ratio for filled and mixed regions remains constant over one order of magnitude in \ha\ and FUV surface brightness, while the shells and clear shells exhibit a wider range of values of almost two orders of magnitude. We derive dust masses and dust temperatures for each \hii\ region fitting the individual SEDs with dust models proposed in the literature. The derived dust mass range is between $10^{2}-10^{4}\msun$ and the cold dust temperature spans $\rm T_{cold}\sim12-27\,K$. The spherical geometrical model proposed for the \ha\ clear shells is confirmed by the emission profile obtained from the observations and is used to infer the electron density within the envelope: the typical electron density is $0.7\pm0.3$\,cm$^{-3}$, while filled regions can reach values two to five times higher.
}
   {}

   \maketitle
%

\section{Introduction}\label{sec:intr}

The interstellar regions of hydrogen ionised by massive stars are normally called 
\hii\ regions. The classical view of an \hii\ region is a sphere of ionised gas whose radius 
is obtained by the ba\-lan\-ce between the number of ionisation  and recombination processes occurring in 
the gas. In a general picture the \hii\ region components are the central ionising stars, a bulk of ionised 
gas which can be mixed with interstellar dust, and a photodissociation region (PDR) surrounding the ionised 
gas cloud and tracing the boundaries between the \hii\ region and the molecular cloud \citep[e.g.][]{Osterbrock:2006p498}. 

The properties of \hii\ regions can be described by the nature of the stellar population which ionises the gas and the physical 
conditions of the interstellar medium (ISM) where the stars are formed. Based on these two aspects we find
 \hii\ regions with a broad range of luminosities, shapes and sizes: from small single-ionised regions to large complexes 
of knots of star formation intertwined with ionised gas filling the gaps between knots. Therefore, the morphology of the regions, 
described by the appearance in \ha\ images, can vary significantly from small concentrated \ha\ distributions to more diffuse 
shell-like structures. 

The physical properties of \hii\ regions have been extensively studied observationally using \ha\ images --in order to obtain luminosity functions, 
mean electron density, and radii \citep[e.g.][]{Kennicutt:1984p117,Shields:1990p799}--, optical spectroscopy --to obtain electron density and temperature, metallicity, and extinction \citep[e.g.][]{McCall:1985p496,Vilchez:1988p505,Oey:2000p121,Oey:2000p124}--, or broad-band images that allows us to study the content of their stellar population \citep[e.g.][]{Grossi:2010p754}. All the different approaches can be combined to gain a better general picture of \hii\ regions. 

Since the recent launch of the {\it Spitzer} telescope, a new wavelength window has been opened to analyse the physical properties of \hii\ regions. Numerous observational studies have been performed in nearby \hii\ regions where the spatial resolution of the {\it Spitzer} data allows us to differentiate between the emission distribution of the Polycyclic Aromatic Hydrocarbons (PAH) features, described by the 8\,\mi\ {\it Spitzer} band and the emission of very small grains (VSG) given by the 24\,\mi\ {\it Spitzer} band 
\citep[among others, for Galactic \hii\ regions,][ for the Large Magellanic Cloud, \citealt{Meixner:2006p786,Churchwell:2006p576}, and for \hii\ regions in M\,33, \citealt{Relano:2009p558,Verley:2009p573,2012ApJ...761....3M}]{Watson:2008p577,2012ApJ...760..149P}. \citet{Churchwell:2006p576} reveals the existence of 322 partial and close ring bubbles in the Milky Way using infrared images from {\it Spitzer} and they argue that the bubbles are formed in general by hot young stars in massive star-forming regions. In M33, \citet{1974A&A....37...33B} presented a catalogue of 369 \hii\ regions over the whole disk of the galaxy and showed the existence of  some ring-like \hii\ regions in the outer parts of the disk. These authors proposed that these regions are late stages in the life of the expanding ionised regions. HI observations of M\,33  \citep[][]{1990A&A...229..362D} reveal the existence of HI holes over the disk of the galaxy: the small (diameters $<$ 500\,pc) HI holes correlate well with OB associations and to a lesser extent with \hii\ regions; however, the large holes (diameters $>$ 500\,pc) show an anti-correlation with \hii\ regions and OB associations. Expanding ionised \ha\ shells have been found in a significant fraction of the \hii\ region population in late-type galaxies \citep{Relano:2005p645}. This can be interpreted in terms of an evolutionary scenario where the precursors of the HI holes would be the expanding ionised \ha\ shells \citep[][and see also \citealt{2012MNRAS.427..625W}]{Relano:2007p790}. 

The high-resolution data from {\it Herschel} instrument \citep{Pilbratt:2010p699} cover a new IR wavelength range that has not been available until now. The combination of data from UV (GALEX) to IR ({\it Herschel}) offers us a unique opportunity to study the SEDs of \hii\ regions with the widest wavelength range up to now. Using new  {\it Herschel} observations, recent studies of a set of Galactic \hii\ regions with shell morphology have been already performed \citep{2012A&A...542A..10A,2012ApJ...760..149P}. Within the Key Project {{\tt HerM33es}}  \citep{Kramer:2010p688}, a set of \hii\ regions has been recently shown in the north part of M\,33 to have an IR emission distribution in the {\it Herschel} bands that clearly follows the shell structure described by the \ha\ emission \citep{Verley:2010p687}. While there is no emission in the 24\,\mi\ band in these regions, cool dust emitting in the 250, 350, and 500\,\mi\ is observed around the \ha\ ring structure. The 24\,\mi\  emission distribution for these objects is very different from the distribution presented in large \hii\ complexes where a spatial correlation between the emission in the 24\,\mi\ band and \ha\ 
emission has been observed  \citep[e.g.][]{Verley:2007p574,Relano:2009p558}. 

Recently, a study on the star-forming regions in the Magellanic Clouds has analysed the relation of the amount of flux at the different wavelengths using SED analysis \citep{Lawton:2010p787}. These authors found that the \hii\ region SEDs peak at 70\,\mi\ and obtained a total-IR (TIR) luminosity from the SED analysis for each \hii\ region. Unfortunately this study does not cover the optical and ultraviolet (UV) part of the spectrum which is crucial to study the amount of stellar radiation that can heat the dust.

We study here the dust emission distribution as well as the dust physical properties in a large sample of \hii\ regions in M\,33 looking for trends with morphology. M\,33, one of the disk galaxies in the Local Group with a significant amount of star formation, is the most suitable object to perform such as multi-wavelength study. The spatial resolution for the bands covering from UV (GALEX) to IR \citep[{\it Herschel}, Key project {{\tt HerM33es}},][]{Kramer:2010p688} allows us to study the interior of the \hii\ regions in this galaxy and to extract the SED of individual objects. The analysis will help us to better understand the interplay between star formation and dust in different \hii\ region types. 

The paper is organised as follows. In Sect.~\ref{sec:data} we present the data we use here, from UV (GALEX) to Far-IR (FIR) ({\it Herschel}). Sect.~\ref{sec:met} is devoted to explain the methodology applied to select and classify the \hii\ region sample and to obtain the photometry of the objects. In Sect.~\ref{sec:SED} we present the SEDs for each \hii\ region and extract conclusions on the SED trends related to the morphology of the objects, and in 
Sect.~\ref{sec:dust} we study the physical properties of the dust for \hii\ regions with different morphology. Sect.~\ref{sec:prof} is devoted to analyse the emission distribution of each band within the individual regions and to derive the electron density for a set of the regions with different morphology. In Sect.~\ref{sec:disc} we discuss the results, and in Sect.~\ref{sec:sum} we present a summary of the main conclusions of this paper.

\section{The data}\label{sec:data}
In this section we describe the multi-wavelength data set that has been compiled for this study. A summary of all the images used here, along with their angular resolutions is given in Table~\ref{tab:data}.

\subsection{Far and near ultraviolet images}

To investigate the continuum UV emission of M\,33, we use the data from GALEX \citep{Martin:2005p691}, in particular the data distributed by \citet{GildePaz:2007p692}. A description of GALEX observations in far--UV (FUV, 1350--1750~\AA) and near--UV (NUV, 1750--2750~\AA) relative to M\,33 and of the data reduction and calibration procedure can be found in \citet{Thilker:2005p693}. The angular resolution of these images are 4\farcs4 and 5\farcs4 for FUV and NUV respectively.  

\begin{table}
\caption{\label{tab:data} Summary of the multi-wavelength set of data.} 
\begin{tabular}{l l l l l}
Telescope & Instrument & Wavelength &  PSF  \\
\hline
          &            &            &    $''$         \\
\hline \hline
GALEX     & FUV        & 1516\,\AA  &  4.4             \\
GALEX     & NUV        & 2267\,\AA  &  5.4             \\
\hline
KPNO      & H$\alpha$  & 6563\,\AA  &  6.6             \\
\hline
\Spi      & IRAC       & 3.6\,$\mu$m&  2.5             \\
\Spi      & IRAC       & 4.5\,$\mu$m&  2.9              \\
\Spi      & IRAC       & 5.8\,$\mu$m&  3.0            \\
\Spi      & IRAC       & 8.0\,$\mu$m&  3.0              \\
\hline
\Spi      & MIPS       & 24\,$\mu$m &  6.3             \\
\Spi      & MIPS       & 70\,$\mu$m &  16.0               \\
\Spi      & MIPS       & 160\,$\mu$m&  40.0              \\
\hline
{\it Herschel}  & PACS       & 100\,$\mu$m&  7.7             \\
{\it Herschel}  & PACS       & 160\,$\mu$m&  11.2             \\
\hline
{\it Herschel}  & SPIRE      & 250\,$\mu$m&  21.2            \\
\hline
\end{tabular}
\end{table}

\subsection{H$\alpha$ images}

To trace the ionised gas, we use the narrow-line H$\alpha$ image of M\,33\ obtained by \citet{Greenawalt:1998p694}. 
The reduction process, using standard IRAF\footnote{IRAF is distributed by the National Optical Astronomy Observatories,
which are operated by the Association of Universities for Research in Astronomy, Inc., under cooperative agreement with the National
Science Foundation.} procedures to subtract the continuum emission, is described in detail in \citet{Hoopes:2000p594}.
The total field of view of the image is $1.75 \times 1.75$~deg$^2$ ($2048 \times 2048$ pixels with a pixel scale of 2\farcs03) with a 6\farcs6 resolution. 

The \ha\ image from the ``Survey of Local Group Galaxies''  \citep{Massey:2006p517} is used here to check for the existence of shells and to revise the morphological classification (see Section~\ref{sec:samp}), as it has a much better angular resolution (0\farcs8) and pixel scale (0\farcs27). Unfortunately, this image is saturated in the central parts of the most luminous \hii\ regions, therefore the photometry has been extracted from the \ha\ image by \citet{Hoopes:2000p594}.

\subsection{Infrared images}

Dust emission can be investigated through the mid-IR (MIR) and FIR data of M\,33\ obtained with the \Spi\ Infrared Array Camera (IRAC) and Multiband Imaging Photometer (MIPS) \citep{Werner:2004p695,Fazio:2004p696,Rieke:2004p697}. The complete set of IRAC (3.6, 4.5, 5.8, and 8.0\,\mi) and MIPS (24, 70, and 160\,$\mu$m) images of M\,33 is described in \citet{Verley:2007p574,Verley:2009p573}: the {\it Mopex} software \citep{Makovoz:2005p698} was used to gather and reduce the Basic Calibrated Data (BCD). We chose a common pixel size equal to 1\farcs2 for all images. 
The images were background subtracted, as explained in \citet{Verley:2007p574}. The spatial resolutions measured on the images are 2\farcs5, 2\farcs9, 3\farcs0, 3\farcs0, for IRAC 3.6, 4.5, 5.8, 8.0\,\mi, respectively; and 6\farcs3, 16\farcs0, and 40\farcs0 for MIPS 24, 70, and 160\,\mi, respectively.
The complete field-of-view observed by \Spi\ is very large and allows us to achieve high redundancy and 
a complete picture of the star-forming disk of M\,33, despite its relatively large extension on the sky. 
The {\it Herschel} observations of M\,33 were carried out in January 2010, covering a field of 1.36 square degrees. PACS (100 and 160
\,$\mu$m) and SPIRE (250, 350, and 500\,$\mu$m) were obtained in
parallel mode with a scanning speed of 20$''$\,s$^{-1}$. The PACS
reduction has been performed using the map-making software Scanamorphos
\citep{2012arXiv1205.2576R} as described in \citet{Boquien:2011p764}. The SPIRE reduction has been done using the
Herschel Data Processing System
\citep[HIPE,][]{2010ASPC..434..139O,2011ASPC..442..347O} and the maps
were created using a "naive" mapping projection
\citep{Verley:2010p687,Boquien:2011p764,2012A&A...543A..74X}. The spatial resolution of the  {\it Herschel}  data are: 7\farcs7 and 11\farcs2 for PACS 100\,\mi\ and 160\,\mi\ and 21\farcs2, 27\farcs2, and 46\farcs0 for 250\,\mi, 350\,\mi, and 500\,\mi, respectively. Due to the significant improvement in spatial resolution of the PACS images, we use here the PACS 100\,\mi\ and 160\,\mi\ images rather than the MIPS 70\,\mi\ and 160\,\mi.

\section{Methodology}\label{sec:met}

In this section we explain how the sample of \hii\ regions is selected and 
how we perform the photometry that allow us to obtain the SED for each 
object.  

\subsection{Sample of \hii\ regions}\label{sec:samp}

We visually select a sample of \hii\ regions and classify them fulfilling the following criteria: 
{\it filled regions} are objects showing a compact knot of emission, {\it mixed regions} 
are those presenting several compact knots and filamentary structures joining the different knots, 
and {\it shells} are regions showing arcs in the form of a shell. We add another classification for 
the special case where we see complete and closed shells, these objects are called {\it clear shells}. 
We use the \ha\ image from  \citet{Hoopes:2000p594} to perform the 
selection of the regions and the morphological classification. In a further step, we check for the classification 
with the high-resolution \ha\ image of \citet{Massey:2006p517}. From the 119 selected \hii\ regions, 
9 are filled, 47 are mixed,  36 are shell, and 27 are clear shell \hii\ regions.

  \begin{figure*}
   \centering
  \includegraphics[width=0.7\textwidth]{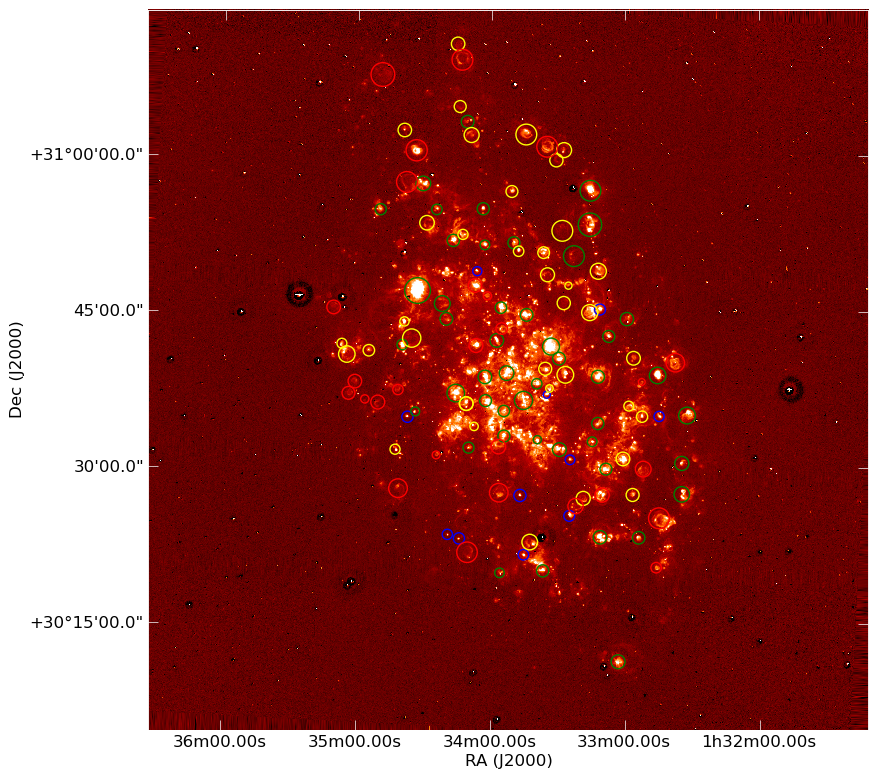}   
   \caption{Location of the \hii\ region sample on the continuum-subtracted \ha\ image of M\,33 from \citet{Hoopes:2000p594}. The radii of the regions  correspond to the aperture radii used to obtain the photometry. Colour code is as follows: filled (blue), mixed (green), shell (yellow) and clear shell (red) regions.}
              \label{hiireg}%
    \end{figure*}
    
\begin{figure*}
   \centering
  \includegraphics[width=0.7\textwidth]{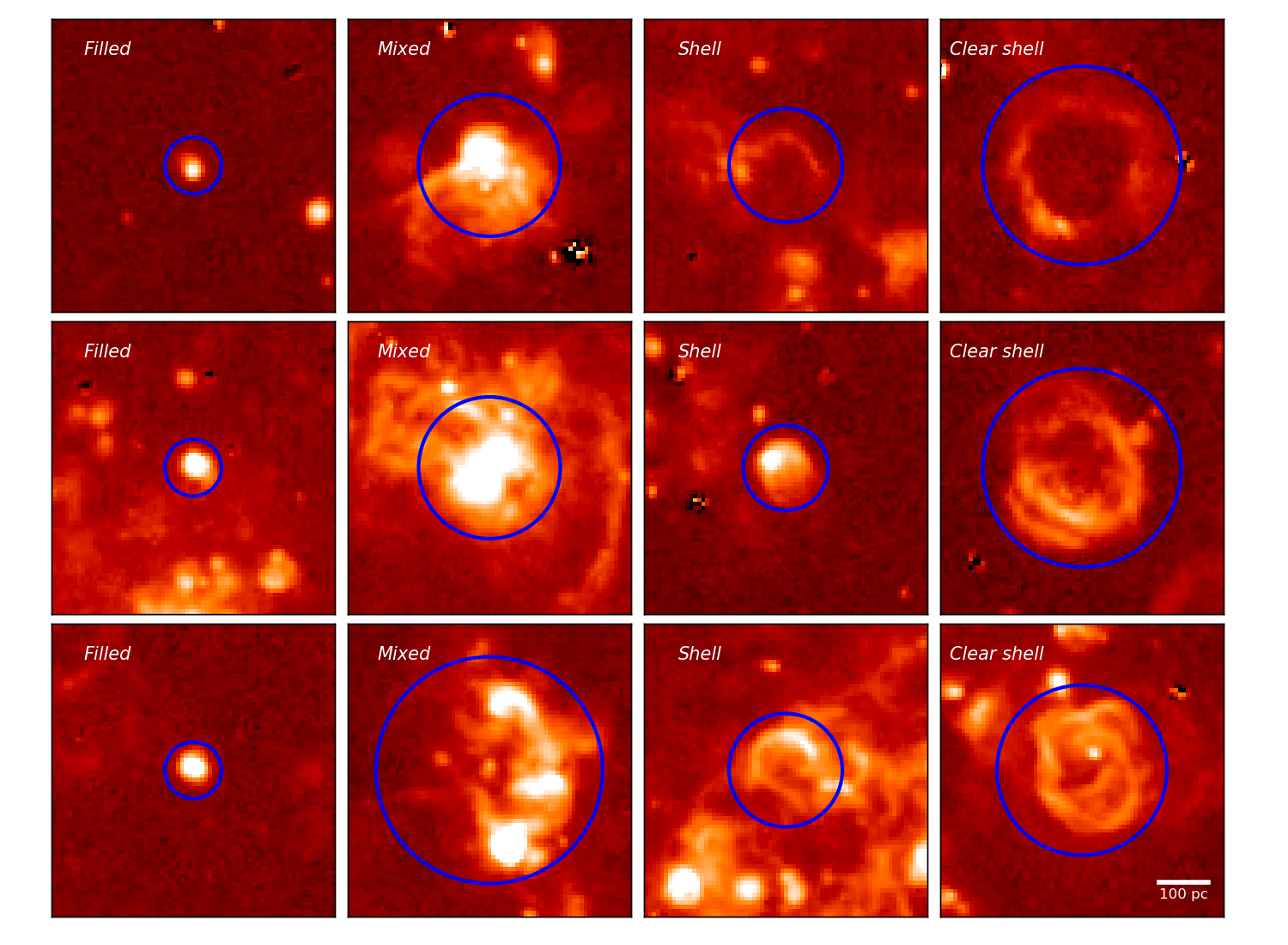}   
   \caption{Examples of \hii\ regions for each classification (see Section~\ref{sec:samp} for details on how the classification was performed). The circles correspond to the aperture used to obtain the photometry. The aperture radii are given in column 4 of Table~B.1.}
              \label{hiireginset}%
    \end{figure*}

In Fig.~\ref{hiireg} we show the continuum-subtracted \ha\ image and the location of our 
\hii\ region sample. A colour code was used to show the different morphological classes: blue, green, yellow, 
and red stand for filled, mixed, shell, and clear shell \hii\ regions, respectively. 
An example of \hii\ regions for each morphology can be seen in Fig.~\ref{hiireginset}, and the WCS coordinates, aperture size and classification are presented in Table~B.1.

\begin{figure*}
   \centering
  \includegraphics[width=0.45\textwidth]{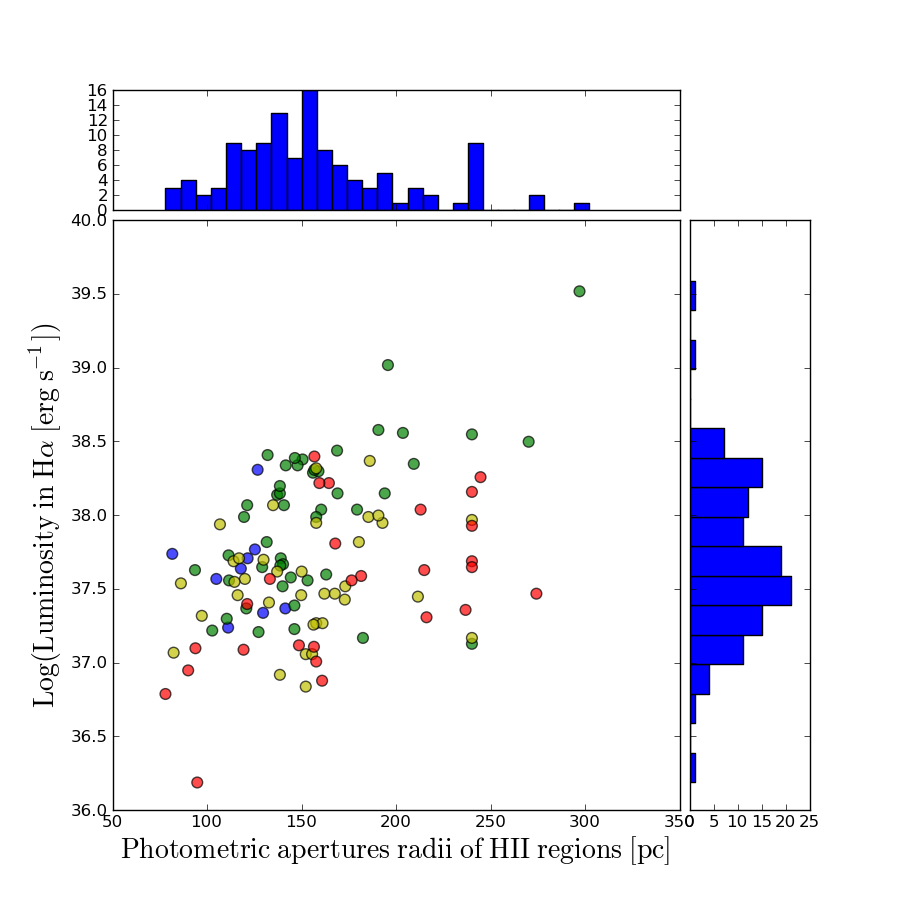} 
  \includegraphics[width=0.45\textwidth]{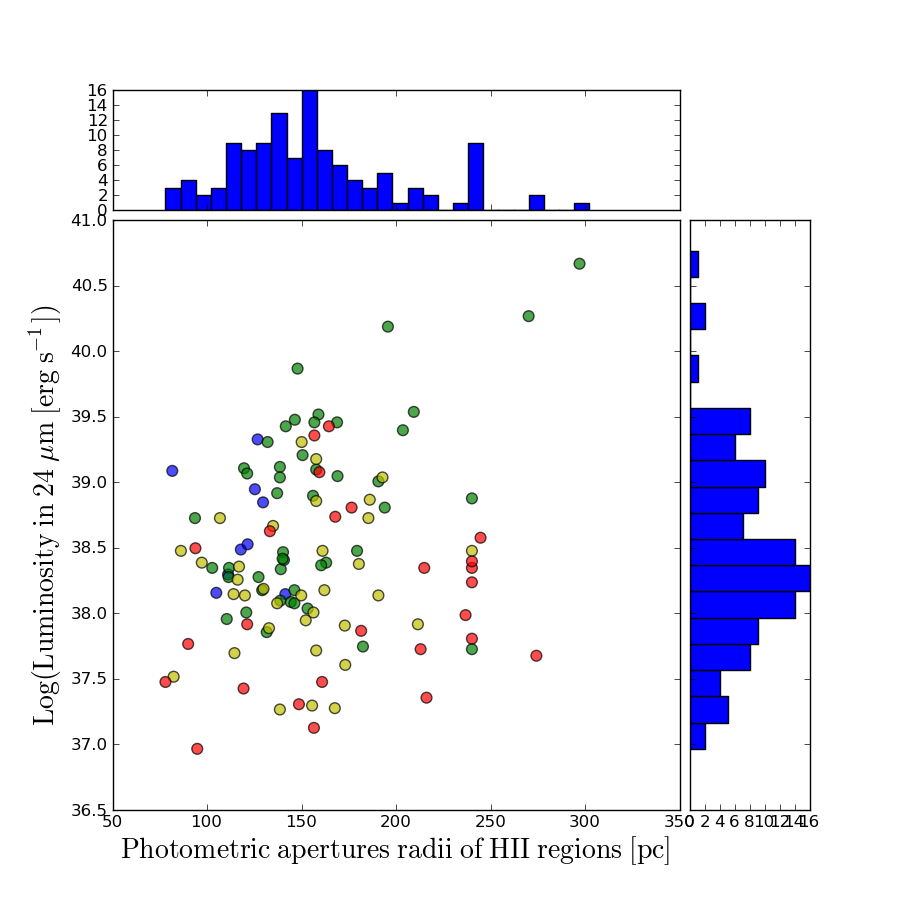} 
   \caption{Histograms of the \ha\ (left) and 24\,\mi\ (right) luminosities of our \hii\ region sample and the aperture sizes used to perform the photometry.}
              \label{ha_24hist}%
    \end{figure*}

The \hii\ region sample is by no means complete, as we chose a set of objects isolated enough to distinguish 
mor\-pho\-lo\-gy. Therefore, we do not attempt to derive any results based on the completeness of the  
\hii\ region population of the galaxy, we rather infer conclusions from the comparison of the SED behaviour 
of \hii\ regions with different \ha\ morphologies. 
Our sample of \hii\ regions presents common objects with previous M\,33 source catalogues given in the literature. 
Using a tolerance of 120\,pc, the 119 sources of the present study overlap with 45 \hii\ regions in \citet{1999PASP..111..685H}, 7 star clusters in \citet{2001A&A...366..498C}, 16 star clusters in \citet{Grossi:2010p754}, 67 sources selected at 24\,\mi\  \citet{Verley:2007p574} and 38 Giant Molecular Clouds in \citet{2012A&A...542A.108G}. The main location difference with \citet{1999PASP..111..685H} and \citet{Verley:2007p574} is that the present study have more objects towards the outskirts of the galaxy as we are selecting isolated \hii\ regions for which clear morphological classification can be carried out.
The star clusters in common with \citet{Grossi:2010p754} show ages between 1.5 and 15\,Myr. In Fig.~\ref{ha_24hist} we show the \ha\ (left) and 24\,\mi\ (right) luminosity distribution of our sample. The luminosities spans a range of more than 2 orders of magnitude in \ha\ and 24\,\mi\ bands. The covered luminosity range of our sample is typical of \hii\ regions of spiral galaxies \citep[e.g.][]{1996A&A...307..735R}.

\begin{figure}
   \centering
  \includegraphics[width=\columnwidth]{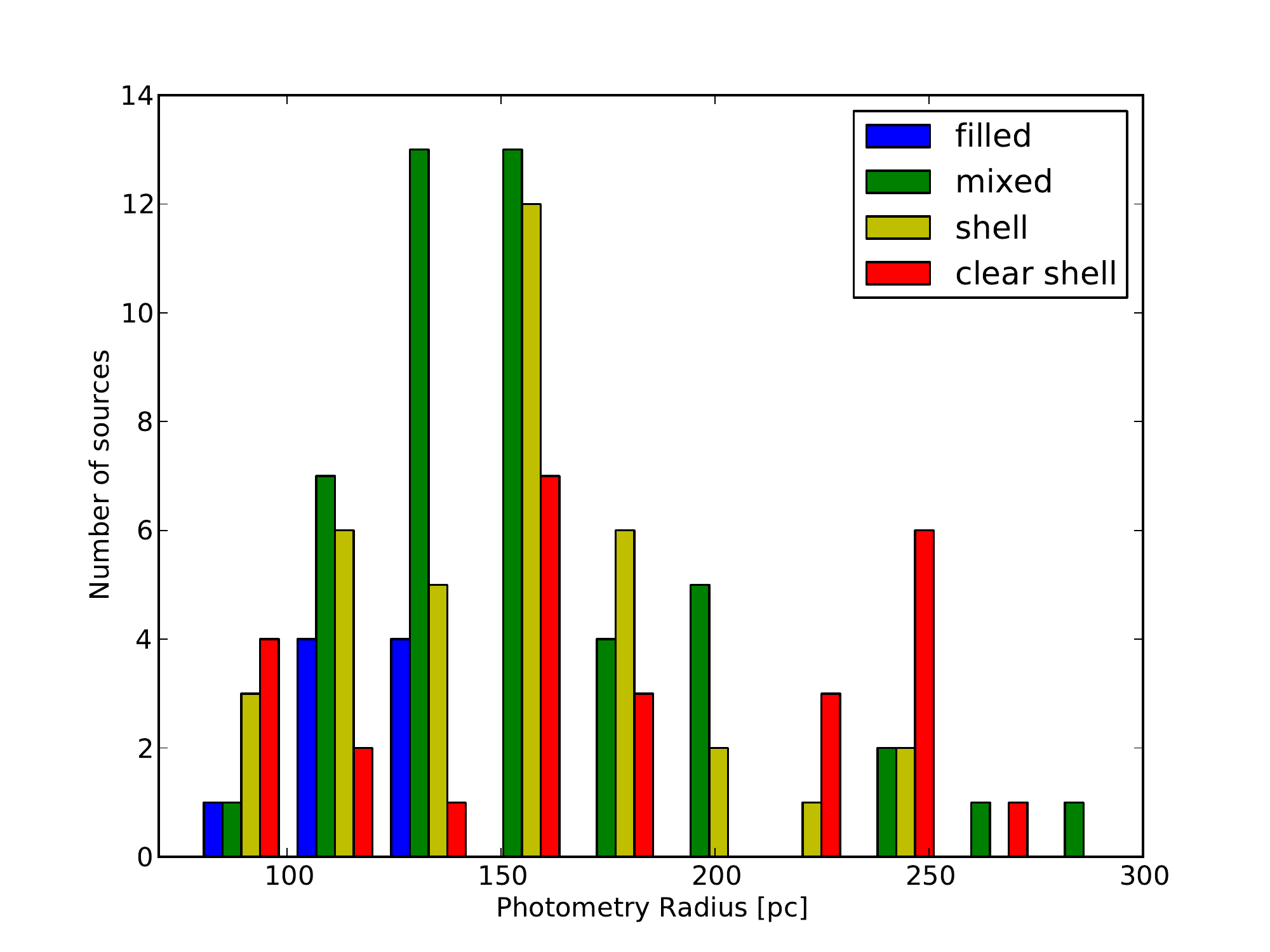} 
   \caption{Histogram of the \hii\ region photometric aperture radii of the classified object sample in M\,33. The photometric aperture was defined using the \ha\ image from \citet{Hoopes:2000p594} (see Sect.~\ref{sec:phot} for more details).}
              \label{histo}%
    \end{figure}

We defined the photometric apertures for each \hii\ region using the \ha\ image from \citet{Hoopes:2000p594} in order to include the total emission of the region. We also compared these apertures with the 24\,\mi\ image to ensure that the emission in this band was also included in the selected aperture. It is important to note that the selection of the \hii\ regions have been made in a visual way, choosing those that are isolated and have clear morphology. Therefore, the photometric aperture size is in some way arbitrary and defines what we think is an \hii\ region showing one of the morphological types analysed in this study. The selected photometric apertures should not be confused with the actual sizes of the \hii\ regions.
A histogram of the aperture radii of the 
classified \hii\ regions is shown in Fig.~\ref{histo}. There is a relation between the photometric aperture size and 
the morphology of the regions: while most filled regions have radii smaller than  
$\sim$150\,pc, the mixed are larger, followed by the shells with radii up to 250\,pc. 
Clear shell regions have much larger radii, up to 270\,pc. The large mixed \hii\ region 
with 280\,pc radii is the largest star-forming region in M\,33, NGC~604. 

There is a relation between the location of the regions on the galaxy disk and the morphological classification. 
Most of the clear shells are seen in the outer parts of the galaxy (see e.g. the north and west outer parts of Fig.~\ref{hiireg}).
However, in order to check whether we are biased by the crowding effects near the centre of the galaxy while defining our sample, we created a set of 10 fake shells with different radii and luminosities in our \ha\ image and redid the morphological classification. We were able to recover only 1 of the fake inserted shells. This simple exercise 
shows that our selection has been done to create a clear defined classification to study the trends of the SED with the 
morphology, rather than to attempt a study of the complete \hii\ region population of M\,33.  

\subsection{Photometry}
\label{sec:phot}
Due to the spatial resolutions and pixel scales of the different images, we perform some technical steps before obtaining the photometry of the regions in each band. We first subtract the sky level in each image when this task was not originally done by the instrument pipelines. Since the angular resolution of the SPIRE 350 and 500\,$\mu$m maps are 27\farcs2 and 46\farcs0, respectively, we decide to discard them and use the SPIRE 250\,$\mu$m as our reference map, degrading all the other maps of our set to a resolution of 21\farcs2, the SPIRE 250\,\mi\ spatial resolution. Keeping a final resolution to better than 21\farcs2 is important in this project since we want to be able to disentangle the structure of shells, which would disappear if we degrade our maps to a worse resolution. We also register our set of images to the SPIRE 250\,$\mu$m image, with a final pixel size of 6.0$''$. Although we still could perform an analysis of the SEDs without the 250\,$\mu$m data, the flux at this band is necessary in order to estimate the dust mass and dust temperature for the individual \hii\ regions (see Section~\ref{sec:dust}).

The photometry has been performed with the IRAF task {\tt phot}. The total flux within the visually defined aperture of 
each source is measured in all the bands. In order to eliminate the contribution of the diffuse medium to the measured fluxes, we subtract a local background value for each region. The local background is defined as the mode value of the pixels within a ring whose inner radii is located 5 
pixels away from the circular aperture and with a 5 pixels width. The mode is obtained after rejecting all the pixels 
with values higher than 2 times the standard deviation value within the sky ring. We choose different widths and inner radii to define the sky annulus and find differences in the sky values of $\sim$5-20\%.  The regions with absolute fluxes lower than their errors are assigned an upper limit of 3 times the estimated uncertainty in the flux. Those regions with negative fluxes showing absolute values higher than the corresponding errors are discarded from the study. In Table~B.2 we show the fluxes for each band together with the corresponding errors.

\section{Spectral Energy Distributions}\label{sec:SED}

\begin{figure*}
   \centering
  \includegraphics[width=0.45\textwidth]{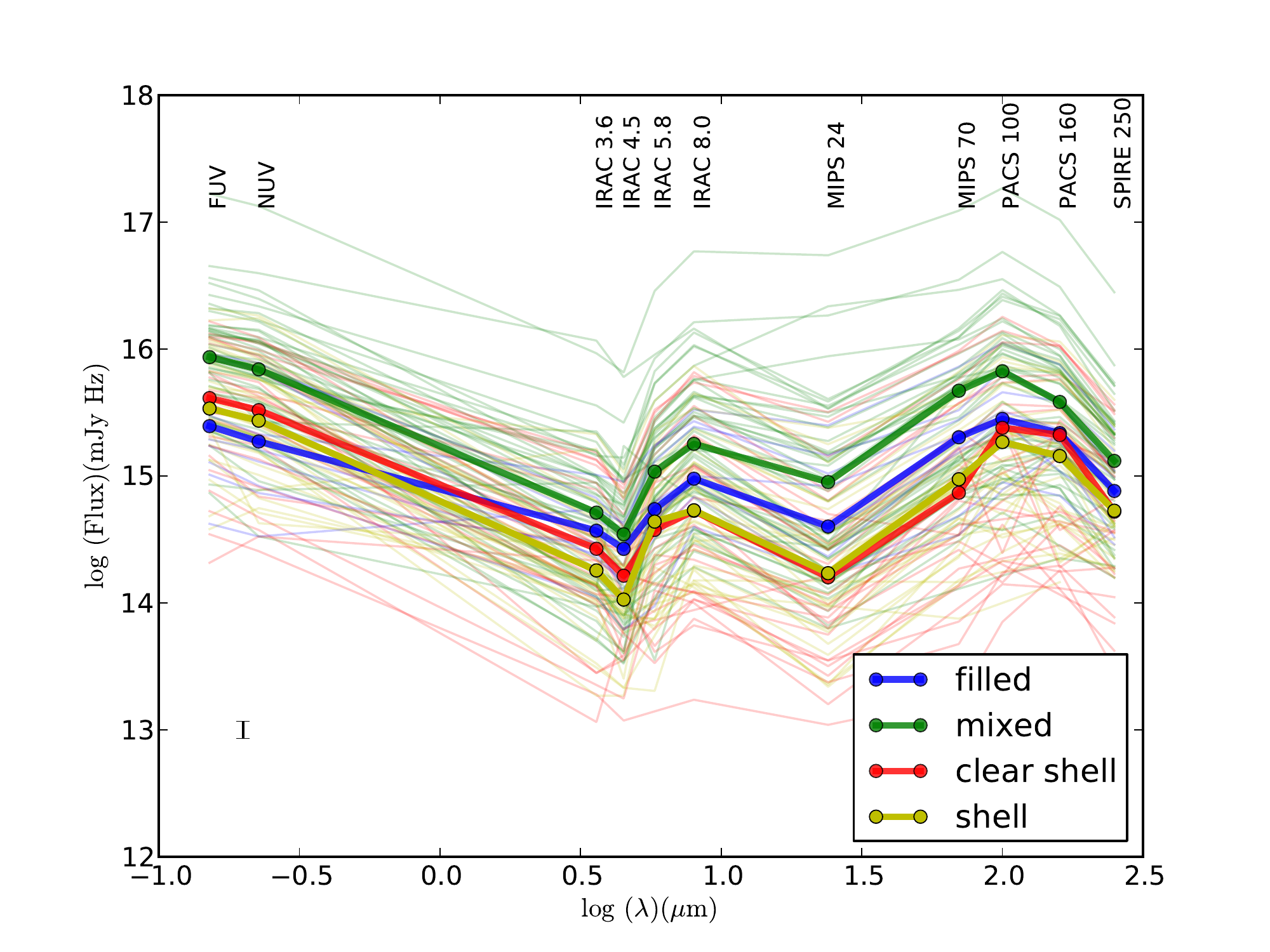}   
    \includegraphics[width=0.45\textwidth]{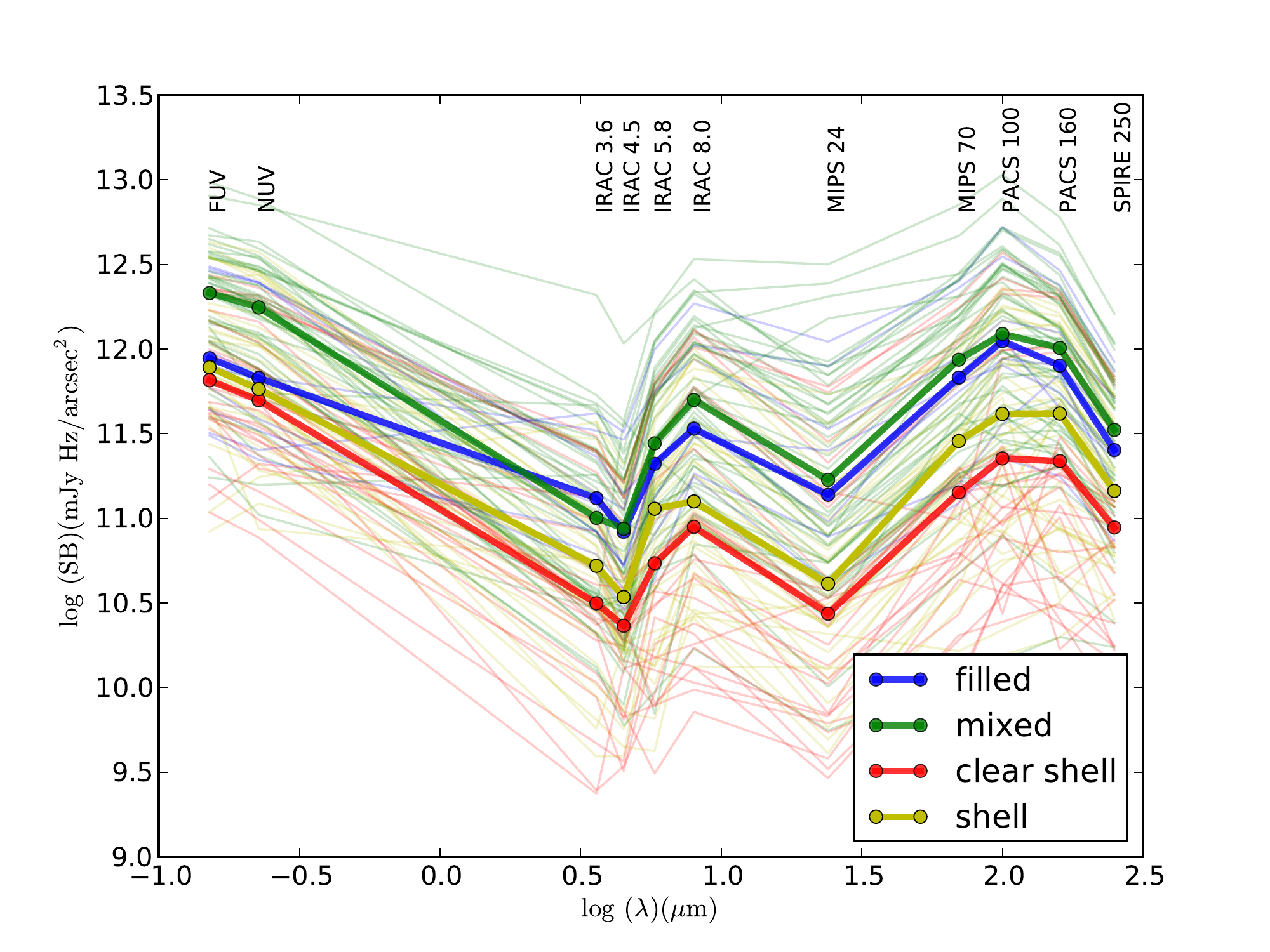}   
   \caption{SED for our set of \hii\ regions. Left panel: SEDs derived using the fluxes of the regions, a median value for the fluxes uncertainties is  shown in the lower left corner. Right panel: Surface brightness is used to obtain the SEDs. The thick lines correspond to the SEDs obtained using the median values at each band for all the \hii\ regions in each morphological sample.}
              \label{sednonorm}%
    \end{figure*}

We obtain the SEDs for each \hii\ region in our sample. The result is shown in Fig.~\ref{sednonorm}, in left panel we represent the total flux versus the wavelength, and in right panel we show the surface brightness (SB). The regions corresponding to each morphology are colour coded: blue, green, yellow, and red correspond to filled, mixed, shell, and clear
shell, respectively. The thicker lines correspond to the SEDs obtained using the median values at each band for all the \hii\ regions in each morphological sample. When calculating the median value at each band, the upper limit fluxes (see Section~\ref{sec:phot}) were discarded. 

Several trends can be seen in these figures showing the different behaviour between the filled-mixed \hii\ regions and the shell-like objects. The \hii\ regions classified as mixed are the most luminous in all bands, which reflects that these \hii\ regions are formed by several knots of \ha\ emission and correspond to large \hii\ region complexes (see left panel of Fig.~\ref{sednonorm}). However, in the right panel of 
Fig.~\ref{sednonorm} we see that the filled and mixed regions are the ones with higher SB. Interestingly, the IR flux for filled, shells, and clear shells are very similar but the SB of the shells and clear shells is lower than SB of filled and mixed regions. This could be interpreted by a pure geometrical argument due to the fact that shells and clear shells cover a larger area than filled regions. 
As the fluxes between the filled, shell and clear shell regions is very similar (left panel of Fig.~\ref{sednonorm}), but the SBs change (filled and mixed regions show similar SB, while the shells and clear shells show lower SB than the filled and mixed ones, see right panel of Fig.~\ref{sednonorm}), we suggest that the filled regions could be the previous stages of the shell and clear shell objects. The explanation is the following: the total flux would be conserved in all the regions but when the region ages and expands the SB lowers (as it is happening for the shells and clear shells). Mixed regions would also fit in this picture as they are typically formed by several filled regions: their total flux should be higher than the filled regions, but the SB should be the same as the filled regions. 

Another interesting point that the SB SEDs show is the low SB at 24\,\mi\ for shells and clear shells (right panel of Fig.~\ref{sednonorm}): the mixed and filled regions show higher emission of hot dust than the shells and clear shells,
which may be due to the proximity of the dust to the power sources in the filled and mixed regions. Besides, the slightly steeper
slopes for the shells and clear shells between 24\,\mi\  and 70\,\mi\ implies that the relative fraction of cold and hot dust could be higher
for shells and clear shells than for filled and mixed regions.

\begin{figure*}
   \centering
  \includegraphics[width=0.45\textwidth]{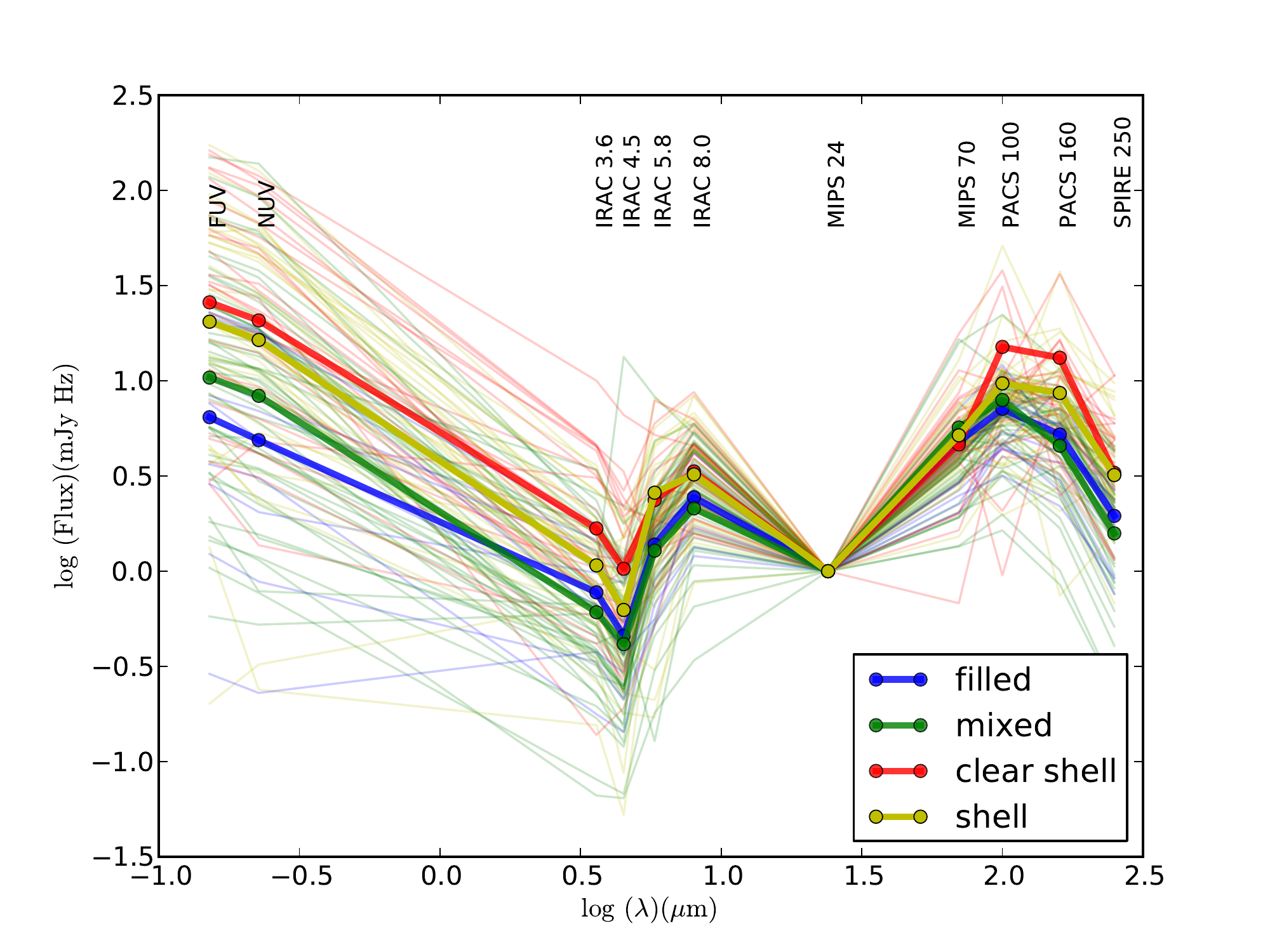}   
  \includegraphics[width=0.45\textwidth]{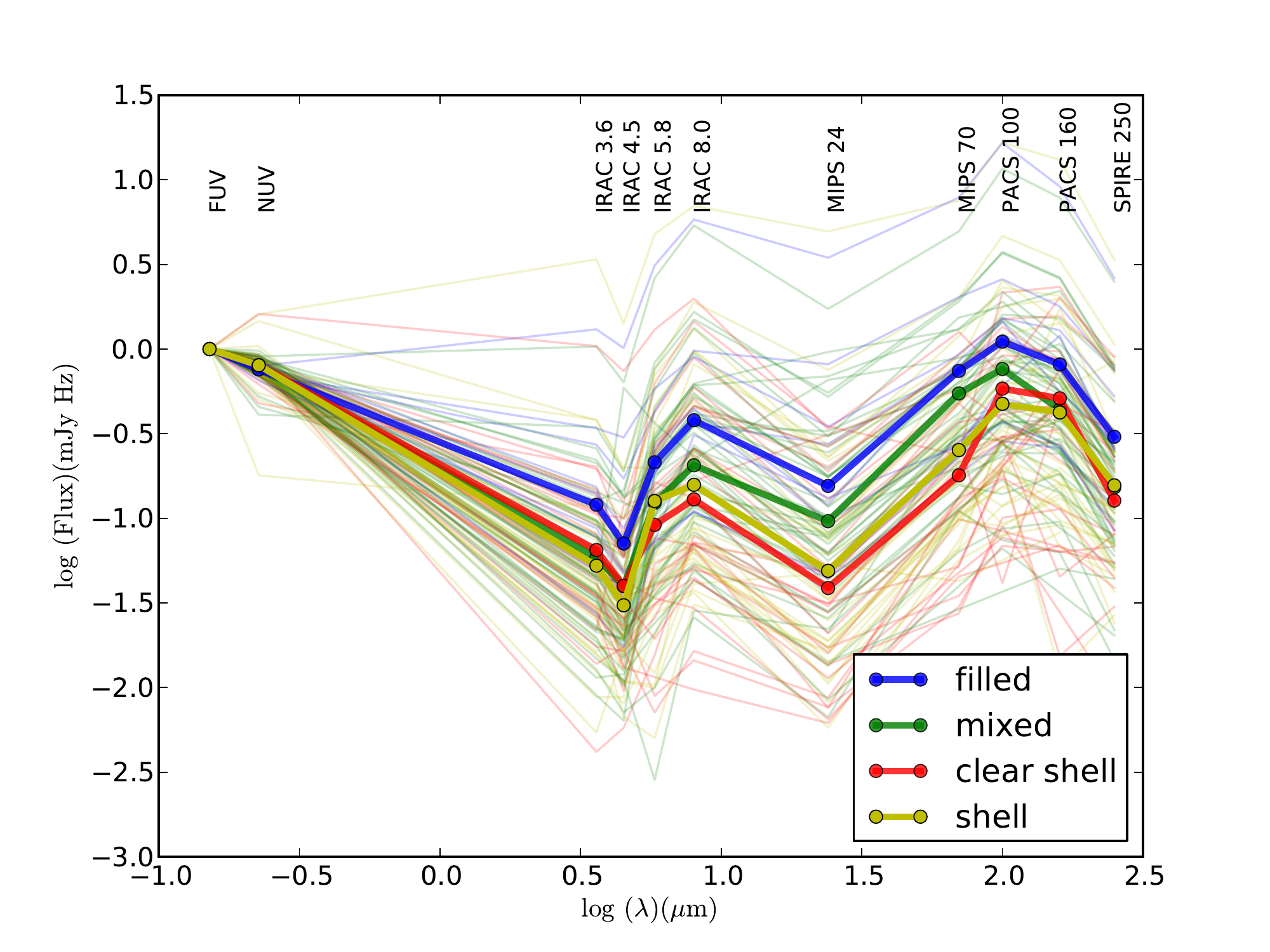}   
   \caption{SED for our set of \hii\ regions normalised to the emission in the 24\,\mi\ band from {\it Spitzer} (left panel) and to the FUV emission from GALEX (right panel). The normalisation emphasises the IR part of the SED showing the different behaviour for filled and mixed regions and shell and clear shells objects. The thick lines correspond to the SEDs obtained using the median values at each band for all the \hii\ regions in each morphological sample.}
              \label{sed24norm}%
    \end{figure*}

The SED trends in the IR part of the spectrum are emphasised when we normalise the SEDs to the 24\,\mi\ fluxes (left panel of Fig.~\ref{sed24norm}). Filled and mixed regions follow the same pattern and have the similar normalised fluxes in the MIPS, PACS and SPIRE bands, while shells and clear shells have in general higher fluxes for these bands. In the right panel of Fig.~\ref{sed24norm} we show the SED normalised to the FUV flux from GALEX. Filled regions show the highest FIR fluxes in this normalisation. This shows that in the filled regions the dust is so close to the central stars that it is very efficiently heated, while shells and clear shells
present less fluxes in this normalisation because the dust is in general distributed further away from the central stars.
Also, in right panel of Fig.~\ref{sed24norm} we see that the FIR peak for shells and clear shells seems to be located towards longer wavelengths, indicating that the dust is colder for this type of object. In a sample of 16 Galactic \hii\ regions \citet{2012ApJ...760..149P} found that the SED peak of the \hii\ regions is located at $\sim$70\,\mi, while the SEDs obtained with larger apertures including the PDR peak at $\sim$160\,\mi. Indeed the SEDs of shells and clear shells might include a higher fraction of PDR, and therefore shifting the peak towards  longer wavelengths. In a study of the SEDs for a set of \hii\ regions in the Magellanic Clouds \citet{Lawton:2010p787} also concluded that most of the SEDs peaks around 70\,\mi. Here we show that the peak of the IR SEDs is closer to the 100\,\mi\ band than to the 70\,\mi\ one. The behaviour of the SED in the IR part of the spectrum will be studied in more detail in section~\ref{dustcoltemp}.

\section{Dust physical properties}\label{sec:dust}

\defcitealias{Draine:2007p588}{DL07} 
In this section we apply models from \citet{Draine:2007p588} (hereafter \citetalias{Draine:2007p588}) to 
study the contribution of the interstellar radiation field (ISRF) to the heating of dust for each \hii\ region 
type and to estimate the dust mass for each individual \hii\ region.

\subsection{Analysis of the stellar radiation field}

The ratio of the surface brightness in different IR bands may bring information about the dust properties. We devote this 
section to study the properties of the dust for our objects and to investigate possible relations between the dust 
properties and the region morphologies. To compare the emission of the dust in the IR bands we subtract 
the stellar emission in the 8\,\mi\ and 24\,\mi\ bands. We use the 3.6\,\mi\ image and the prescription given by 
\citet{Helou:2004p788} to obtain a pure dust ({\it non-stellar}) emission at 8\,\mi\ ($F_\nu^{\rm\it ns}(8\mi)$) and at 24\,\mi\ ($F_\nu^{\rm\it ns}(24\,\mi)$).  
\begin{eqnarray}
F_\nu^{\rm\it ns}(8\mi) & = & F_\nu(8\mi) - 0.232 F_\nu(3.6\mi)  ; \\
F_\nu^{\rm\it ns}(24\,\mi) & = & F_\nu(24\,\mi) - 0.032 F_\nu(3.6\mi). 
\end{eqnarray}
\citetalias{Draine:2007p588} suggest three ratios to describe the properties of the dust (see Eq. 3, 4, and 5).
In these equations, 1) $P_{8}$ corresponds to the emission of the PAHs and 2) $P_{24}$ traces the thermal hot dust. 
These quantities are normalised to the $\nu F_\nu(71\mi) + \nu F_\nu(160\mi)$, which is a 
proxy of the total dust luminosity in high intensity radiation fields. 
3) The ratio $R_{71}$ is sensitive to the temperature of the dust grains dominating the 
FIR, and therefore is an indicator of the intensity of the starlight heating the dust. 
\begin{eqnarray}
P_{8} & = & \frac{\nu F_\nu^{\rm ns}(8\mi)}{\nu F_\nu(71\mi) + 
              \nu F_\nu(160\mi)} \\
P_{24} & = & \frac{\nu F_\nu^{\rm ns}(24\,\mi)}{\nu F_\nu(71\mi) + 
              \nu F_\nu(160\mi)}  \\
R_{71} & = & \frac{\nu F_\nu(71\mi)}{\nu F_\nu(160\mi)} 
\end{eqnarray}
\begin{figure*}
   \centering 
  \includegraphics[width=\columnwidth]{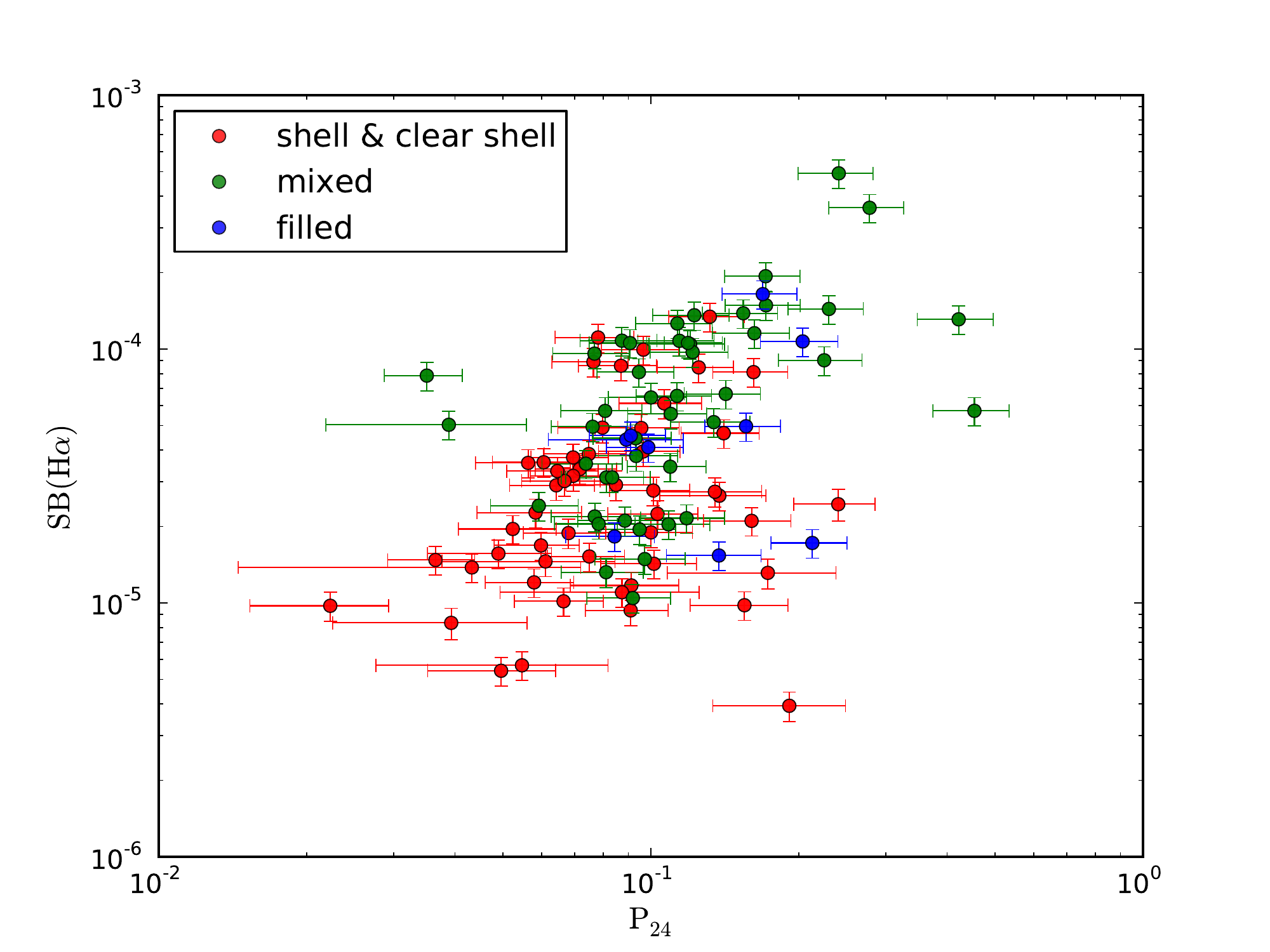}  
   \includegraphics[width=\columnwidth]{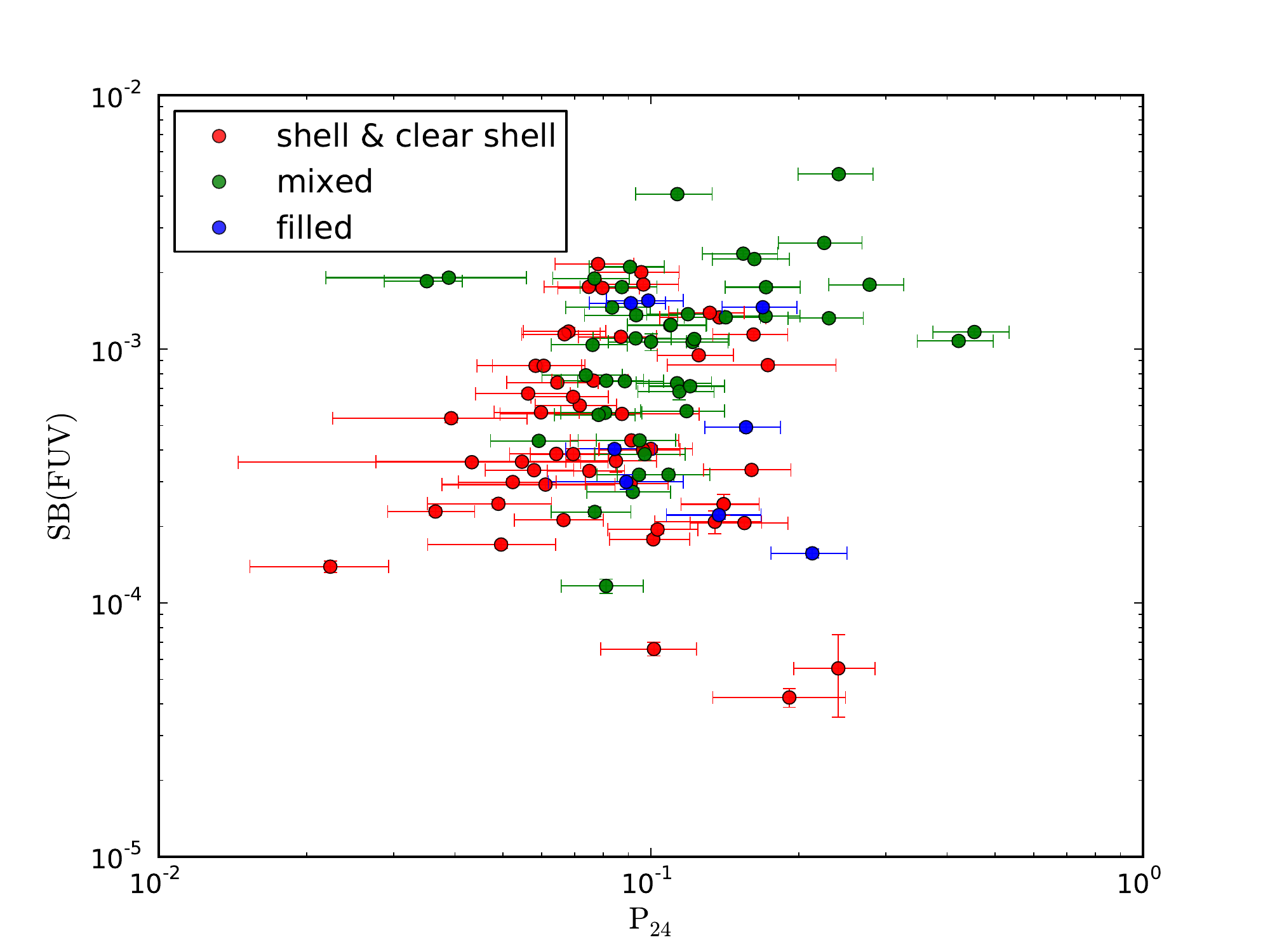}   
   \caption{\ha\ (left) and FUV (right) surface brightness as a function of $P_{24}$. Colour code: {\it red} corresponds to shells and clear shells, {\it blue} to filled and {\it green} to mixed regions.}
              \label{P24vradint}%
\end{figure*}
In Fig.~\ref{P24vradint} we show \ha\ (left) and FUV (right) SB versus $P_{24}$ for \hii\ regions with different morphology.  In these figures blue represents filled regions, green mixed ones, and red stands for shells and clear shells. $P_{24}$ seems to correlate with the SB(\ha) better than with SB(FUV). This seems plausible as $P_{24}$ traces the hot dust, which is related to a younger stellar population and therefore with \ha\ emission. However an extinction effect which is higher for FUV than for \ha\ might be also affecting both diagrams. 
The mixed regions (green points) occupy the top-right part of the diagram in left panel, corresponding to higher SB(\ha) and 
higher $P_{24}$, while the shells and clear shells (red points) show low values of SB(\ha)  
and $P_{24}$. 

In order to better understand how the dust behaves in regions 
of different morphology we have applied \citetalias{Draine:2007p588}'s models to our set of \hii\ regions. \citetalias{Draine:2007p588}'s models reproduce the emission of the dust exposed to a range of stellar radiation fields. 
The models separate the emission contribution of the dust in the diffuse ISM, heated by a general diffuse radiation field\footnote{The values given here for the radiation field are scaled to the interstellar radiation field for 
the solar neighbourhood estimated by \citet{Mathis:1983p593}, $u_{\nu}^{(MMP83)}$. The specific energy density of the 
star is taken to be $u_{\nu}=Uu_{\nu}^{(MMP83)}$ with $U$ a dimensionless scaling factor.} (U$_{\rm min}$), from the emission of the dust close to young massive stars, where the stellar radiation field (U$_{\rm max}$)
is much more intense. The term $(1-\gamma)$ is the fraction of the dust mass exposed 
to a diffuse interstellar radiation field, U$_{\rm min}$, while $\gamma$ would be the corresponding fraction for dust mass exposed 
to U$_{\rm max}$. The models are parameterized by q$_{\rm\sc PAH}$, the fraction of dust mass in the form of PAHs, along with U$_{\rm min}$, U$_{\rm max}$, and $\gamma$.

\begin{figure*}
 \includegraphics[width=\textwidth]{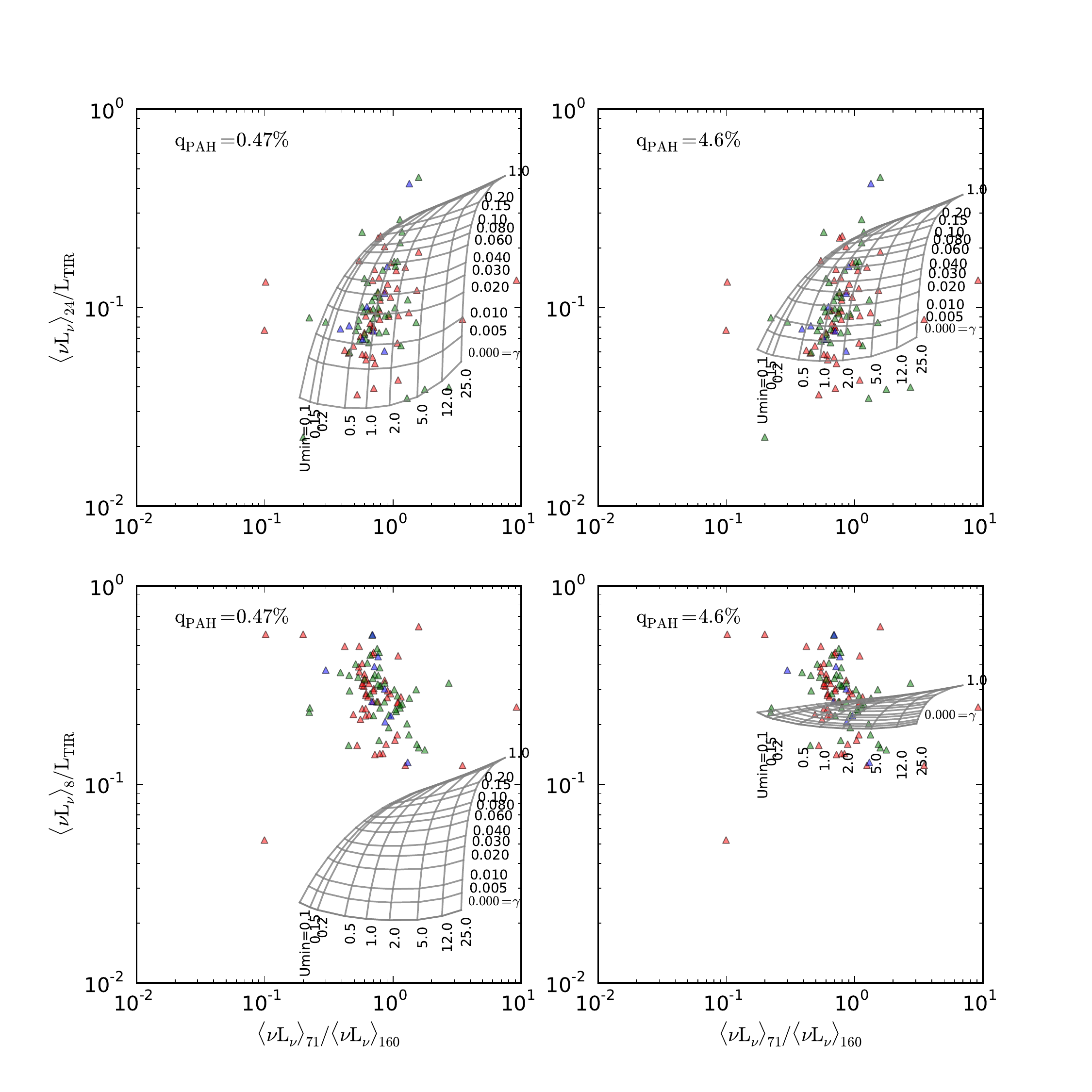}
 \caption{$P_{24}$ (top) and $P_{8}$ (bottom) versus $R_{71}$ for \citetalias{Draine:2007p588} dust models with q$_{\rm\sc PAH}$= 0.47\% (left column) and 4.6\% (right column). A value of U$_{\rm max}=10^{6}U$ has been considered for all models. Colour code: {\it red} corresponds to shells and clear shells, {\it blue} to filled and {\it green} to mixed regions.}
\label{DLmod}
\end{figure*}

In Fig.~\ref{DLmod} we show $P_{24}$ (top) and $P_{8}$ (bottom) versus $R_{71}$ with models from \citetalias{Draine:2007p588} over-plotted. 
The right column shows the models with a fraction of dust mass in PAHs, q$_{\rm\sc PAH}$, of 4.6\%, while in the left column the fraction is 0.47\%. The fraction q$_{\rm\sc PAH}$=4.6\% represents a low limit for our data: most of the regions show a higher PAH fraction than 4.6\% (see bottom-right panel of Fig.~\ref{DLmod}). Since there are no \citetalias{Draine:2007p588} models with q$_{\rm\sc PAH}>$4.6\%, a comparison with our data for higher PAH fraction cannot be carried out and we will proceed the comparison using a fraction of dust mass of 4.6\% (see section~\ref{sec:dustmass}).

In Fig.~\ref{DLmod} (top-right) we show $P_{24}$ versus $R_{71}$ for q$_{\rm\sc PAH}$=4.6\%. The plot shows that a typical fraction of a maximum of $\sim$6\% (i. e. values of $\gamma$ less than 0.06) of the radiation field heating the dust in the shells and clear shells corresponds to young stars. There are some regions showing higher fraction of radiation field heating the dust due to young stars (values of $\gamma$ higher than 0.10), these are mixed or filled regions.  
This would shows the effect of the relative location of the dust and stars in heating the dust: in filled regions the ionised gas and the dust is very close to the stars that can heat the dust, while in the shells the gas and the dust are located further away from the stars. In the last case the stars are less able to heat the dust located at further distances, while in the former case the stars are more efficient to heat the dust located nearby. Therefore, due to the dust-gas and stars configuration, a low fraction of radiation field coming from young stars is expected to heat the dust in regions with shell morphology. Besides, the shell and clear shell regions have larger radii and therefore they extend in general over a larger area in the disk than filled regions (see Fig.~\ref{histo}), and therefore they can be affected by a higher fraction of general diffuse radiation field. 

The mixed regions and the majority of filled regions can be well described by a relative constrained value of U$_{\rm min}\sim$0.5-2 (see Fig.~\ref{DLmod}, top-right) showing that the radiation field coming from the diffuse part of the galaxy, corresponding to an older stellar population, is low for these regions. However, for shell and clear shell regions when we move down in the diagram (corresponding to lower values of gamma and therefore higher fractions of radiation field coming from the diffuse medium) we see a spread in the distribution of data. The explanation for this spread for the shell and clear shell regions is that these low luminosity objects are very affected by the conditions of the ISRF in their surroundings. 

We would like to mention here that the comparison of \citetalias{Draine:2007p588} dust models with the observations of our set of \hii\ regions presented here is merely qualitative. 
Our intention is to find general differences in the dust heating mechanisms in each classification that can help us in order to perform a more detailed study for each individual region. Detailed models of individual \hii\ regions with different morphology are in progress (Rela\~no et al. 2013, in preparation).

Using the plots in Fig.~\ref{DLmod} we are also able to constrain the fraction of the total radiation field corresponding to the old stellar population of the galaxy disk. We find an upper limit of U$_{\rm min}$=5. For the case of the radiation field coming from young stars (U$_{\rm max}$) the constraint cannot be applied as the models are degenerated for values higher than U$_{\rm max}\sim10^6-10^7$. The behaviour of the  $P_{24}$ versus $R_{71}$ diagram is expected as M\,33 is a galaxy with a moderate SFR \citep[$\sim$0.5\msun\,yr\me][]{Verley:2009p573}. For a more passive galaxy with a lower SFR we would have data in the lower part of 
Fig.~\ref{DLmod} (top-right) and lower U$_{\rm max}$ would be needed to explain the data, while for a starburst galaxy higher values of U$_{\rm max}$ would be required.  

\subsection{Morphology and dust colour temperature}
\label{dustcoltemp}

\begin{figure*}
   \centering
  \includegraphics[width=\columnwidth]{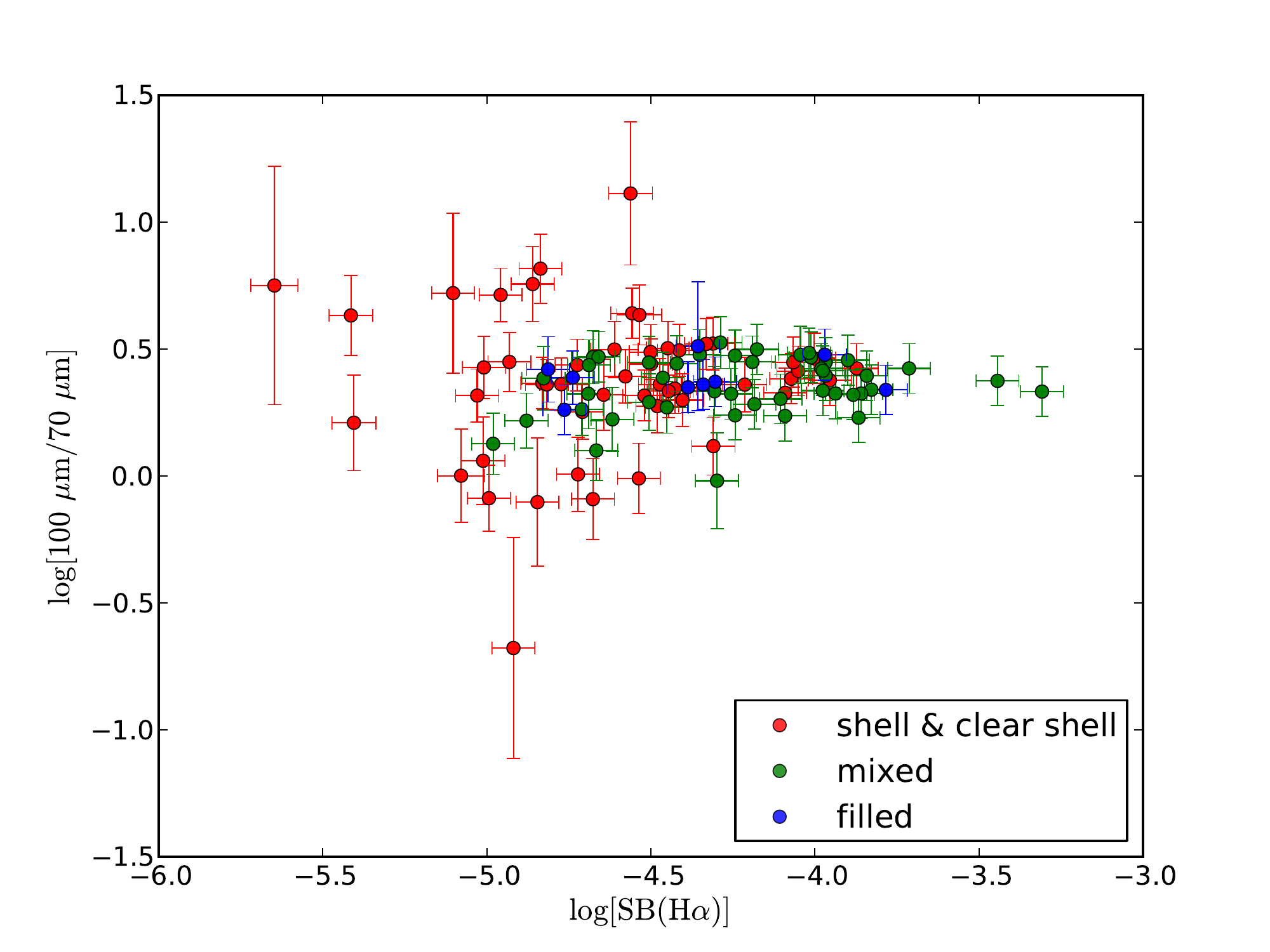}  
   \includegraphics[width=\columnwidth]{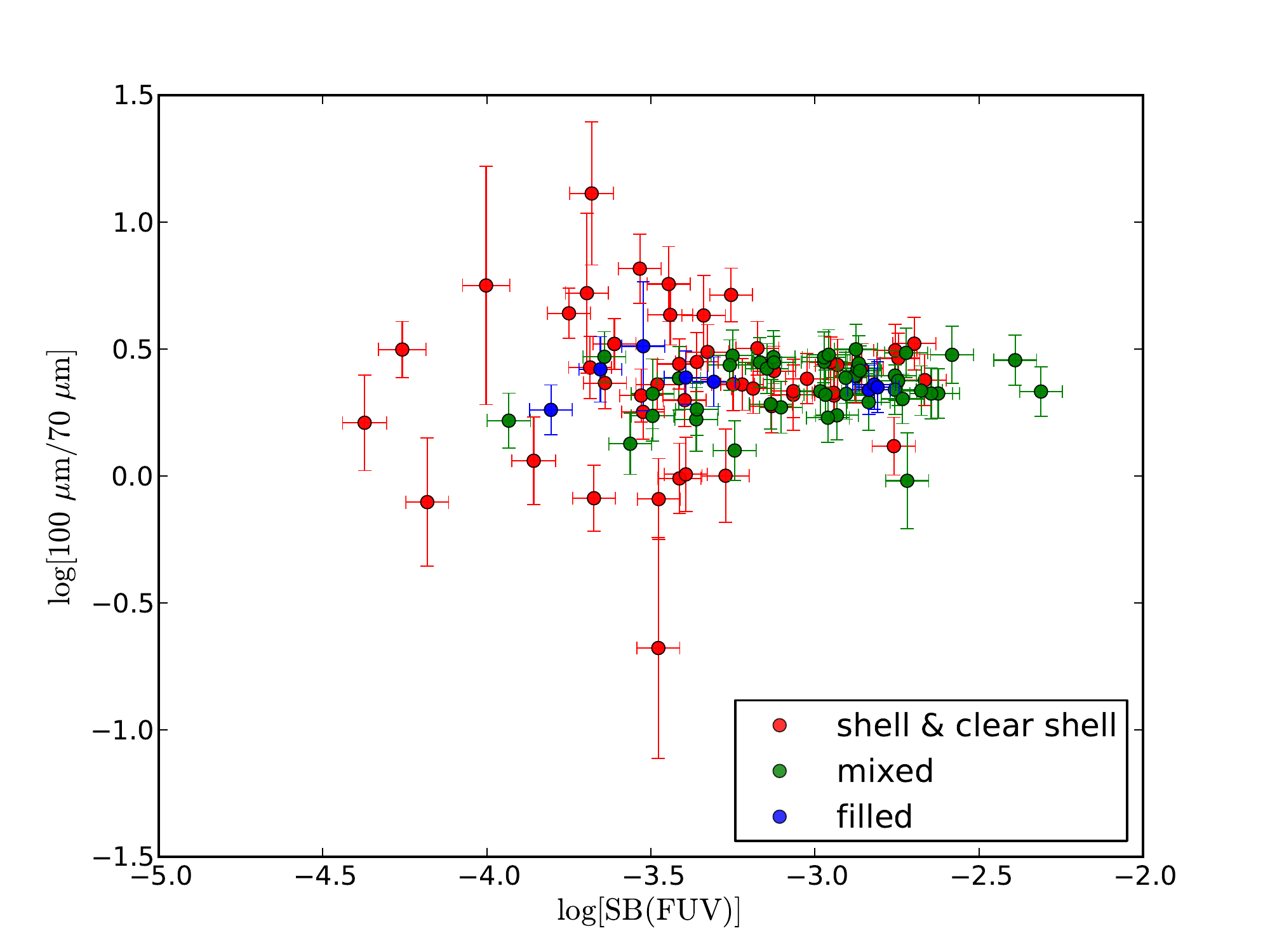}   
   \caption{Left: The 100\,\mi/70\,\mi\ ratio versus the \ha\ surface brightness for the \hii\ region sample. Colour code is the same 
   as in Figs.~\ref{P24vradint} and~\ref{DLmod}. Right: The same as left but versus the FUV surface brightness.}
              \label{F100to70}%
\end{figure*}
The dust temperature can be estimated using the ratio of bands close to the peak of the IR SED. The ratios 100\,\mi/70\,\mi, 160\,\mi/70\,\mi, or 100\,\mi/160\,\mi\ usually trace the temperature of the warm dust emitting from 24\,\mi\ to 160\,\mi, while the cold dust is traced by wavelengths larger than 160\,\mi. With the new window opened by the {\it Herschel} observations the temperature of the cold dust  can be estimated using 250\,\mi/350\,\mi\ and 350\,\mi/500\,\mi\ ratios \citep[e.g.][]{Bendo:2012p789}. In a sample of disk galaxies, \citet{Bendo:2012p789} found that the 70\,\mi/160\,\mi\ ratio for  individual locations across the galaxy disk shows a correlation with \ha\ emission: 70\,\mi/160\,\mi\ ratio is increasing at high \ha\ surface 
brightness. A similar result was shown in \citet{Boquien:2010p690,Boquien:2011p764} for star-forming regions and for individual locations within the disk of M\,33, respectively. This would indicate that at more intense radiation fields the dust would be hotter, which would be the case if the radiation coming from the stars within the \hii\ regions would be the dominant factor to take into account when describing the heating of the warm dust. However, at low \ha\ surface brightness a higher dispersion in the correlation is found, showing that in this regime other mechanisms besides the radiation coming from the stars would be affecting the heating of the warm dust. 

In Fig.~\ref{F100to70} we show the logarithmic 100\,\mi/70\,\mi\ ratio versus the logarithmic \ha\ (left) and FUV (right) surface brightness. In general, the same correlation as the one found by \citet{Bendo:2012p789} and \citet{Boquien:2010p690,Boquien:2011p764} is seen in these figures. However, the logarithmic 100\,\mi/70\,\mi\ ratio remains constant for filled (blue points) and mixed (green) regions over one order of magnitude in surface brightness, and therefore, the warm dust temperature tends to be constant for filled and mixed \hii\ regions. For shells and clear shells the logarithmic 100\,\mi/70\,\mi\ ratio shows a wider range of values of almost two orders of magnitude. This could mean that for filled and mixed objects the dust is so close to the stars within the regions that it is very efficiently heated and reaches a very well defined, narrow range of temperature, independently of the radiation field intensity. For shells and clear shells, other parameters may affect the dust heating mechanism. 
It could probably be the location of the dust relative to the stars or the evolutionary state of the stellar population within the region that may lead to a dispersion in the correlation at low intensity radiation fields.

\subsection{Dust mass}\label{sec:dustmass}

\begin{figure*}
   \centering
  \includegraphics[width=\columnwidth]{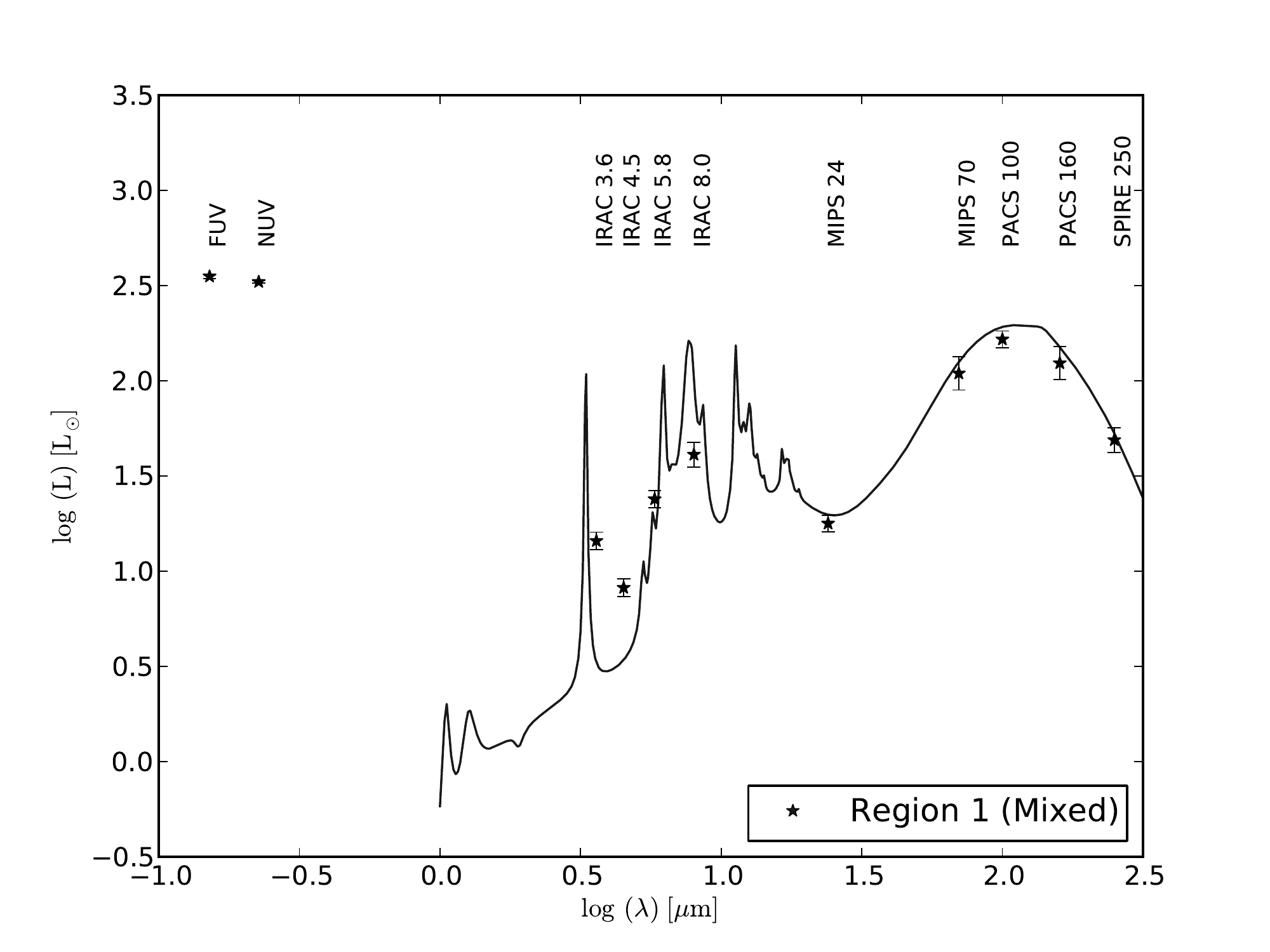}  
   \includegraphics[width=\columnwidth]{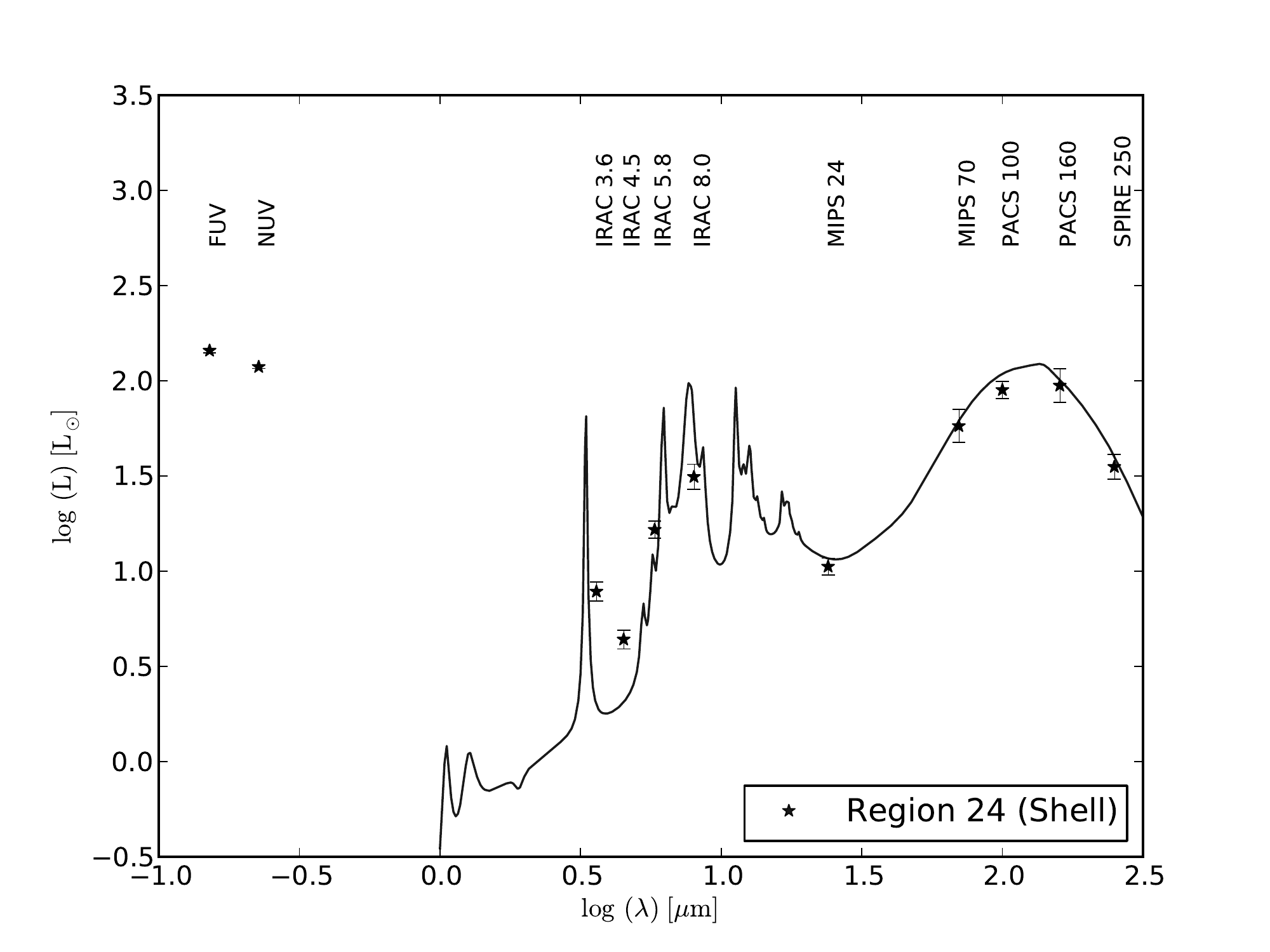}   
  \includegraphics[width=\columnwidth]{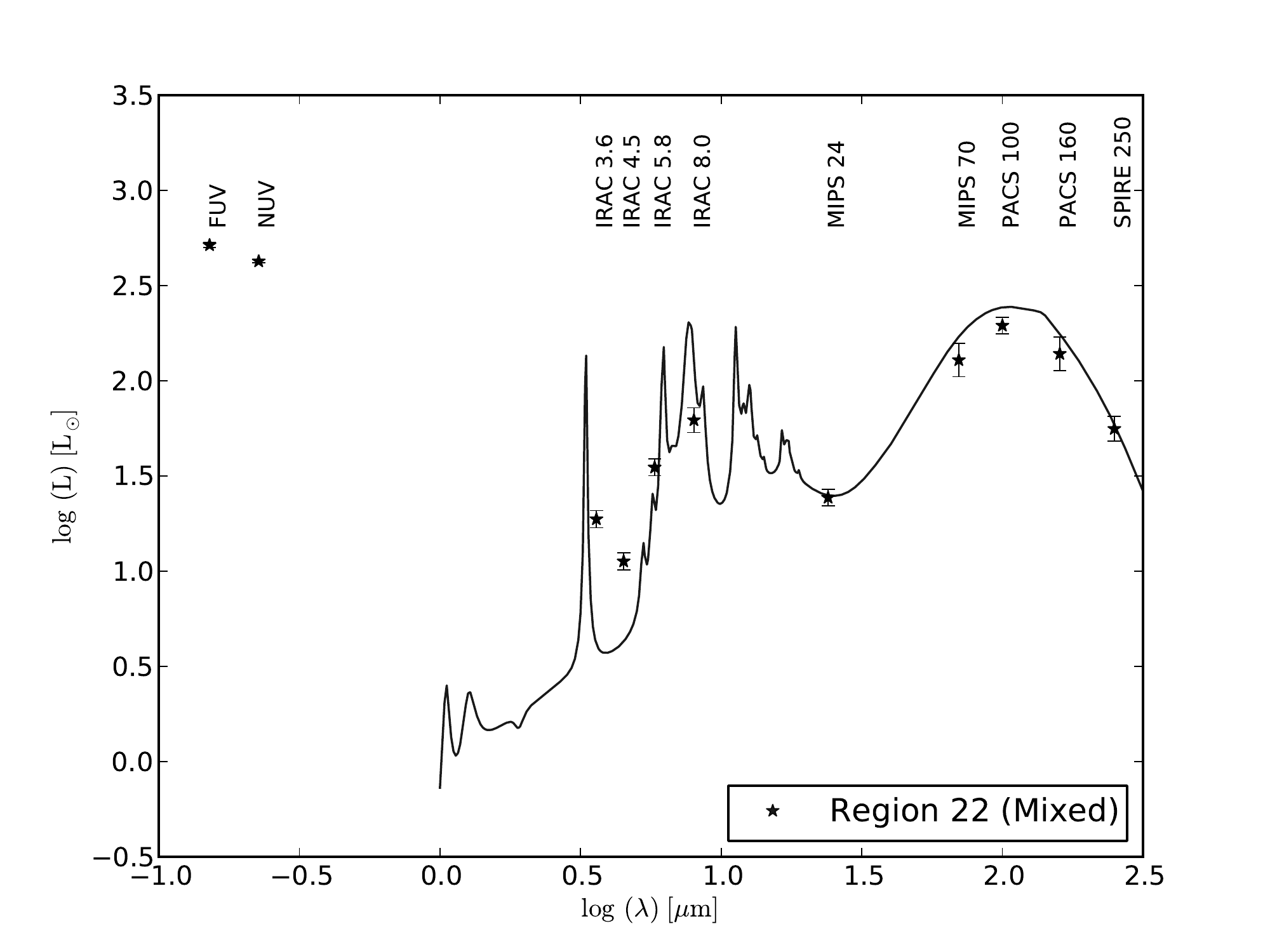}  
   \includegraphics[width=\columnwidth]{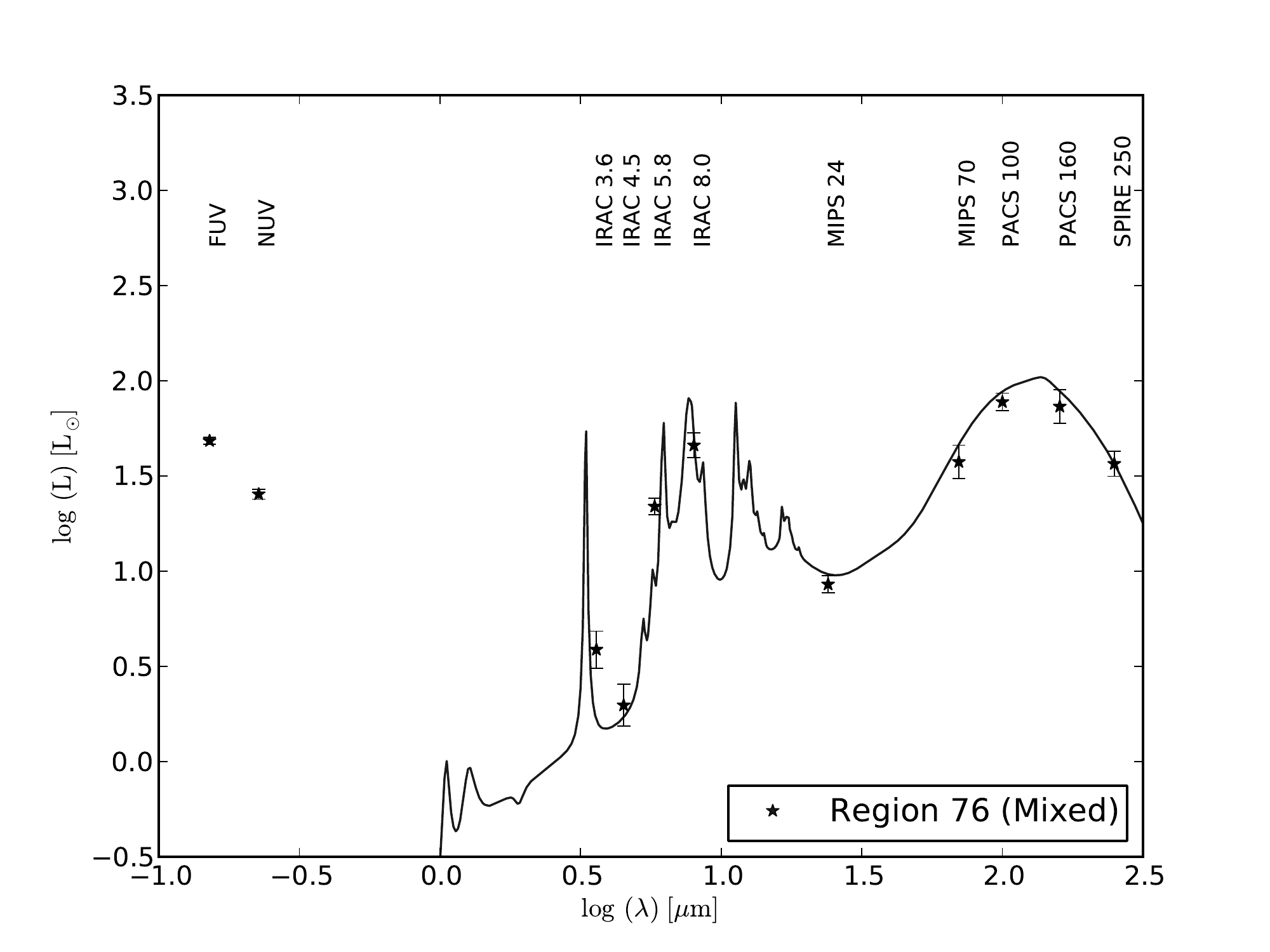}   

   \caption{SED for a sample of \hii\ regions fitted with \citetalias{Draine:2007p588} dust models. Values for the reduced chi-square are: 
   0.84, 0.72, 0.53, 1.07 for \hii\ regions 1, 24, 22, and 76, respectively.}
              \label{DL07fit}%
\end{figure*}

We estimate the dust mass for the \hii\ regions fitting \citetalias{Draine:2007p588} models to the SED of each individual \hii\ region. Since the combination of radiation fields suggested by \citetalias{Draine:2007p588} does not seem to represent well the radiation field heating the dust within the individual \hii\ regions (see Fig.~\ref{DLmod}), we have decided to use a single radiation field U to describe the stellar field of the region. The PAH fraction was fixed to the highest value provided by the models, q$_{\rm\sc PAH}$=4.6\%. We only use bands with wavelengths longer than 8\,\mi, as the PAH features, traced by IRAC 3.6\,\mi-8\,\mi, do not have strong influence in the derivation of the dust mass \citep[see][]{2012ApJ...756..138A}. Since we are interested in deriving the dust mass and compare it with the dust temperature traced by 250\,\mi/160\,\mi\ ratio, we only fitted the \hii\ regions with reliable fluxes in the 160\,\mi\ and 250\,\mi\ bands (see section~\ref{sec:phot}). 

\begin{figure}
   \centering
    \includegraphics[width=\columnwidth]{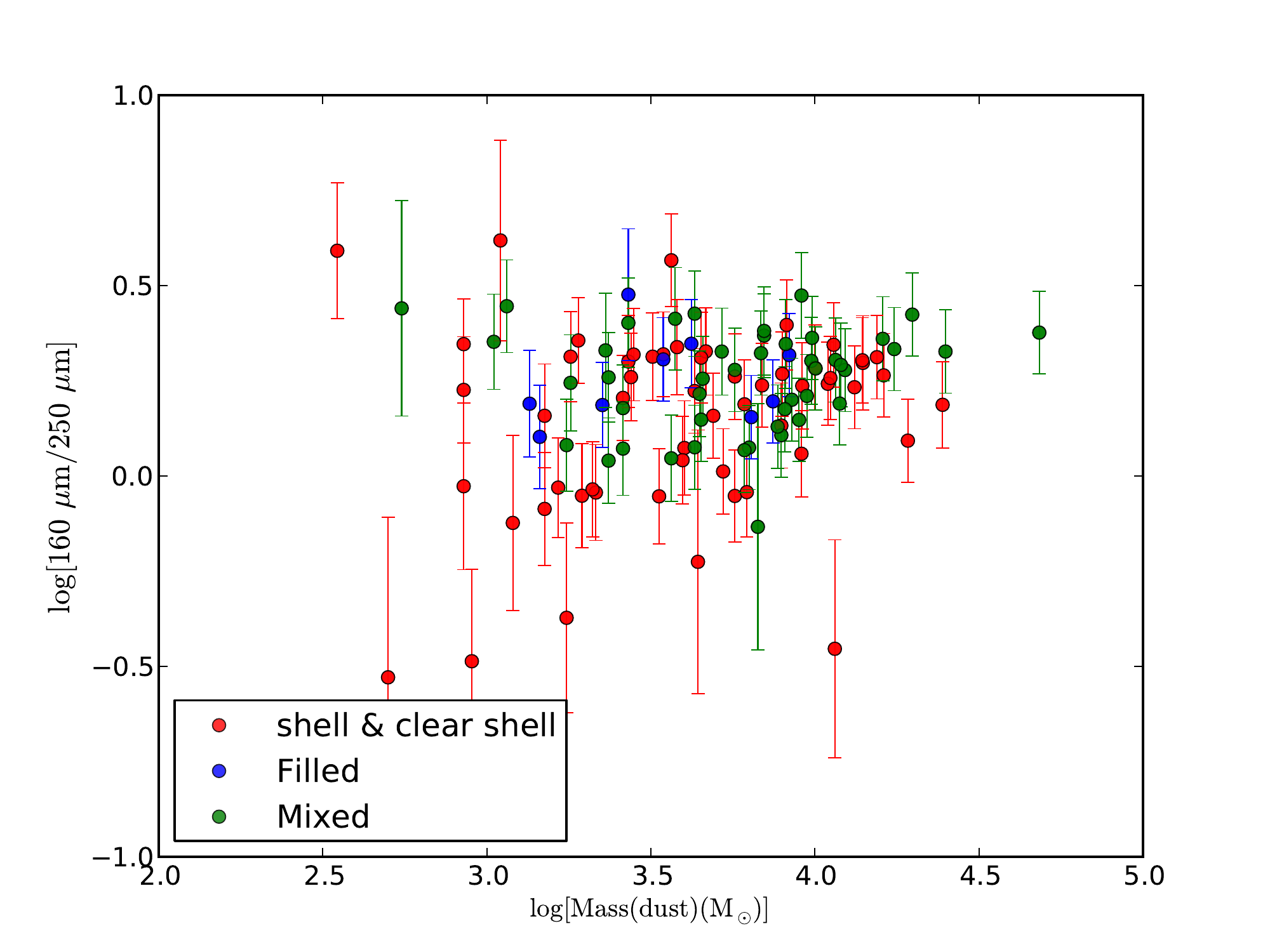}    
   \caption{ Left: 160\mi/250\mi/ ratio versus the dust mass for the \hii\ region sample. Colour code is the same 
   as in Fig.~\ref{P24vradint}.}
              \label{dustmass}%
\end{figure}

In Fig.~\ref{DL07fit} we show some examples of the fit performed to the individual \hii\ regions. In general the models fit relatively well even the IRAC bands that were not included in the fit procedure.  We find dust masses in the range of $10^{2}-10^{4}\msun$ (see Fig.~\ref{dustmass}), which are consistent with dust mass estimates of the most luminous \hii\ regions in other galaxies using \citetalias{Draine:2007p588} models 
\citep[NGC~6822,][]{2010A&A...518L..55G}. The dust masses derived here correspond to the total dust mass included within the aperture chosen to extract the photometry. At the spatial resolution provided by {\it Herschel} data we can not infer whether the dust is completely or partially mixed with the ionised gas within the region.

In Fig.~\ref{dustmass} we show the logarithmic 250\,\mi/160\,\mi\ ratio versus the dust mass for our \hii\ region sample. The logarithmic 250\,\mi/160\,\mi\ ratios are within the range -0.5 to 0.5, which corresponds to a dust temperature range of 10-30\,K \citep[see Eq.\,6 in][]{2012MNRAS.425..763G}. This range of dust temperature agrees with the range of temperatures shown in the map of M33 in \citet{Braine:2010p689} and derived using the 350\,\mi/250\,\mi\ ratio and a modified blackbody fit. The range also agrees with the estimates of the cold dust temperature provided by \citet{2012A&A...543A..74X} for individual locations in M33 and by \citet{2012ApJ...760..149P} for Galactic \hii\ regions using a combination of two modified blackbodies describing the warm and cold dust temperature. Using 250\,\mi/160\,\mi\ as an estimator of the dust temperature we show in Fig.~\ref{dustmass} that the shell and clear shell regions (red data points) tend to have lower dust temperature than the filled and mixed regions (blue and green data points). In Table~B.3 we show the dust mass derived for each \hii\ region with reliable 250\,\mi\ and 160\,\mi\ fluxes. We also present an estimate of the dust temperature provided by the models using the relation given in \citet{2012MNRAS.425..763G}: $\rm T_{cold}(K)=17.5\times U_{min}^{1/6}$, where in our case $\rm U_{min}$ corresponds to the single radiation field, U,  used in the fit. The cold dust temperature is within the range of $\rm T_{cold}\sim12-27\,K$.

\section{Multi-wavelength profiles of H$\alpha$ shells} \label{sec:prof}

We analyse the observed distribution of the multi-wavelength emission along the observed profiles (subsect.~\ref{subsec:profiles_observations}), we propose a three-dimensional model for the clear shells which is validated by the distribution of the \ha\ emission in the envelope (subsect.~\ref{subsec:profiles_models}), and finally, based on this geometrical model, we are able to estimate the electron density (subsect.~\ref{subsec:profiles_ne}) in the envelope of the shells and compare it with the electron density measured in filled regions.

\subsection{Profile analysis} \label{subsec:profiles_observations}
We perform a multi-wavelength study of the emission distribution in the \hii\ regions in each morphological classification. The idea is to look for trends in the spatial distribution of the emission at each wavelength to infer the location of the different gas, dust, and stellar components in the regions. From UV-GALEX to the 250\,\mi\ from {\it Herschel} we obtain profiles into two directions for each \hii\ region: horizontal (East-West) and vertical (North-South), encompassing the centre of the selected \hii\ region. Each profile corresponds to the integration of 4~pixels ($\sim$24\arcsec) width. In Figs.~\ref{prof_clearshell1} and \ref{prof_clearshell3}, we show the emission line profiles of two of the \hii\ regions classified as clear shells and in Figs.~\ref{prof_compact20} and \ref{prof_compact39} of two examples of filled \hii\ regions.

Several trends are clearly seen from the set of analysed profiles and the examples shown here are representative of these trends. The \ha\ profiles of the clear shells show the characteristic double peak of the shell emission when the shell is spatially resolved at the 21\farcs2 resolution of our set of images. The typical sizes of the resolved shells (marked as the spatial separation of the two peaks) are $\sim$300\,pc ($\sim$75\arcsec) while the not-resolved shells \citep[those regions classified as clear shells but not spatially resolved are classified as shells because they show shell structure in the high-resolution \ha\ image of][]{Massey:2006p517} have typical sizes of $\sim$100\,pc ($\sim$25\arcsec). The sizes of the \ha\ shells are in agreement with the vertical scale length of 300\,pc obtained for the ionised gas disk from a Fourier analysis by \citet{2012A&A...539A..67C}. Indeed, \citet{2012A&A...539A..67C} found that the ionised gas lies in a layer thicker than stellar ($\approx$50\,pc) or neutral gaseous ($\approx$100\,pc) disks. The Fourier transform analysis results in one single mean value for the break scale at a given wavelength, but it is probable that the flaring of the ionised gas disk confines the ionised gas in a thinner layer towards the centre of the galaxy and in a thicker layer towards the outskirts. Through a wavelet analysis of the \ha\ map of M\,33, \citet{Tabatabaei:2007p575} found that shells can be as large as 500\,pc. This is in agreement with our study and this could explain why the shells and clear shells, as a mean, could reach larger radii at large galactocentric radius. This confirms the trends, larger \hii\ region radii with respect to the distance from the M\,33 nucleus, shown by \citet{1974A&A....37...33B}.

In the lower panels of Figs.~\ref{prof_clearshell1} and \ref{prof_clearshell3} we show the dust emission distribution in the shells. The emission at all IR bands follows clearly the \ha\ shape of the shell: at 24\,\mi\ and 250\,\mi\ the emission decays in the centre of the shell and it is clearly enhanced at the boundaries (see Fig.~\ref{prof_clearshell3}). The same trend is seen for 70\,\mi, 100\,\mi, and 160\,\mi\ but not as clear as for 24\,\mi\ and 250\,\mi\ emissions. The emission of the old stars (3.6\,\mi\ and 4.5\,\mi) follows in general a different distribution with their maxima generally displaced from the \ha\ maxima (see middle panel of Fig.~\ref{prof_clearshell1}). The emission of the PAH at 8\,\mi\ is marked by the location of the shell boundaries.

The emission distribution at all wavelengths (FUV/NUV and dust emission) for the filled shells follows quite clearly the \ha\ emission. The typical sizes of the filled regions are 20\arcsec, corresponding to the spatial resolution. However, we know that these are filled \hii\ regions and not shells as the classification was checked with the high-resolution \ha\ images of Local Group Galaxies Survey \citep{Massey:2006p517}. The \ha\ emission line profiles show the same functional form as those modelled by \citet{2011ApJ...732..100D} assuming that the radiation pressure is acting on the gas and dust within the \hii\ region. \citet{2011ApJ...732..100D} parameterises the existence of the cavity depending on the value of $Q_{\rm o}n_{\rm rms}$, the number of ionising photons times the electron density of the region. The fact that we already see the cavities in the \ha\ emission line profiles at values of $Q_{o}n_{rms}$ much lower than model predictions shows that the radiation pressure is not the only effect acting here, as it is the case of the Galactic \hii\ region N49 \citep{2011ApJ...732..100D}.

\subsection{Geometrical models of clear shells} \label{subsec:profiles_models}
Most of the shells and clear shells which can be observed in M\,33 appear as circular in projection onto the plan of the sky while the galaxy presents an inclination of 56$^\circ$ \citep{Regan:1994p733}. Using a simple geometrical argument, one can infer that their geometrical shape, in three dimension, is spherical: if this was not the case, one would expect the projection to depend from shell to shell and that statistically, most of the shells would appear
in projection as ellipses or any other form dependent on the particular conditions of the shell and its surrounding ISM.

\begin{figure}
\includegraphics[width=\columnwidth]{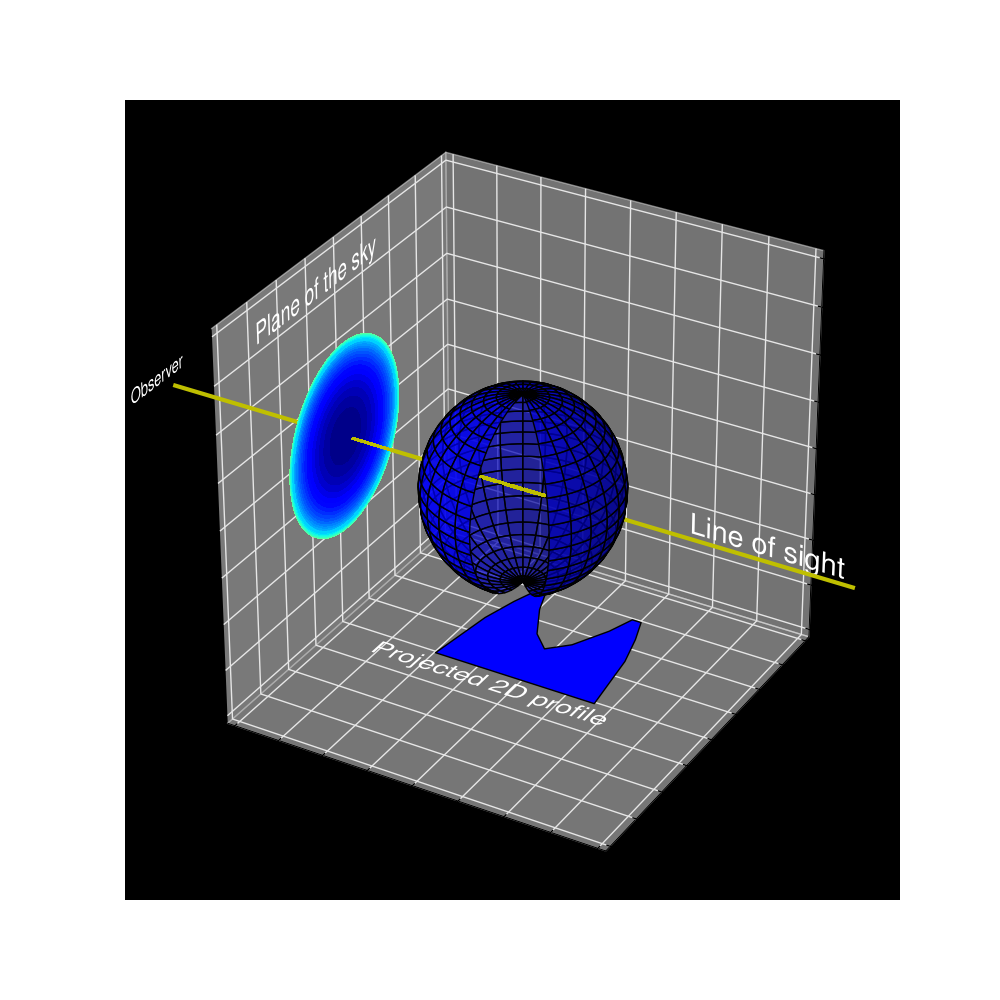}
\caption{Model of one clear shell. The clear shell is modelled by a pure sphere (central blue sphere in the figure; the width of the shell is not represented for sake of clarity). The projection of the shell on the sky will always appear as circular, independently of the galaxy inclination (represented by the circle projected on the plane on the sky, between the observer and the modelled shell). The density profile obtained by collecting the light through a slit placed in front of the circular projection and encompassing its centre is also represented (see the projected 2D profile in the bottom of the figure, placed here also for sake of clarity).} \label{fig:shell3D}
\end{figure}

Following this assumption, we tried to reproduce one of the clearest shells in our sample, lying away from the grand design structures of the galaxy which may add noise to the data by artificially increase or decrease some background emission and break the symmetry. We constructed our model in order to reproduce as closely as possible the Clear shell 1, located in the northern part of the galaxy, at about 6-7\,kpc from the M\,33 centre. The H$\alpha$ horizontal profile of this shell is presented in Fig.~\ref{prof_clearshell1} (upper left panel). The data shows that the maxima of the H$\alpha$ emission are located at a radius of about 120\,pc from the centre of the shell and that the width of the shell is about 60\,pc. Pure geometry (see Fig.~\ref{fig:shell3D}), in the optically thin limit, tells us that the locations of the maxima (about 120\,pc in this example) would trace the inner boundary limit of the shell while the outer boundary is given by the full width of the two horns (a radius of about 180\,pc in the present case).

\begin{figure}
\includegraphics[width=\columnwidth]{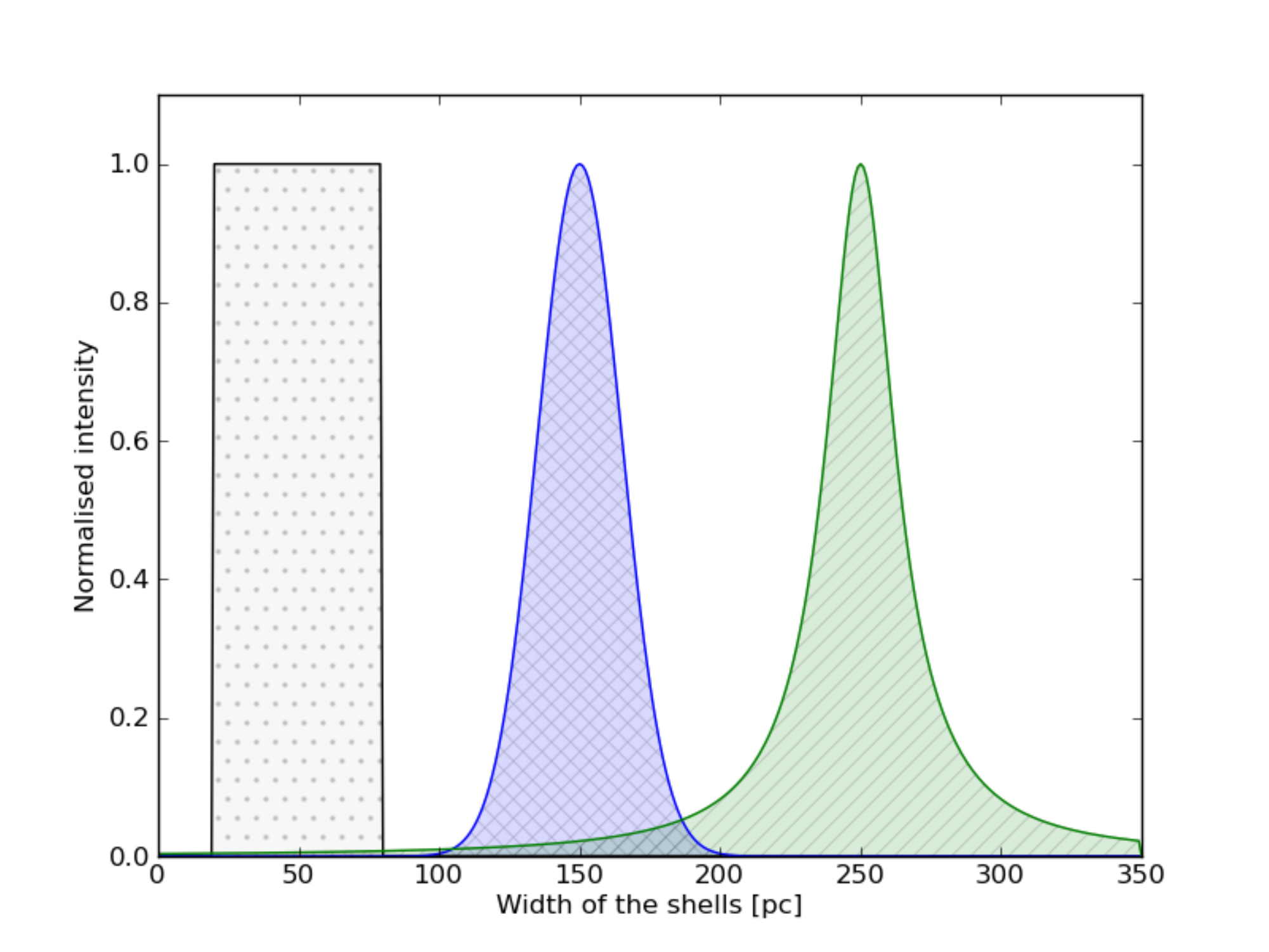}
\caption{Density distribution of the shells. The left (grey) profile is used for the model of constant density. The central (blue) profile follows a Gaussian distribution. The right (green) profile follows a Cauchy-Lorentz distribution. The three profiles are artificially shifted along the abscissa for the clearness of the plot.} \label{fig:profileDensity}
\end{figure}

\begin{figure}
\includegraphics[width=\columnwidth]{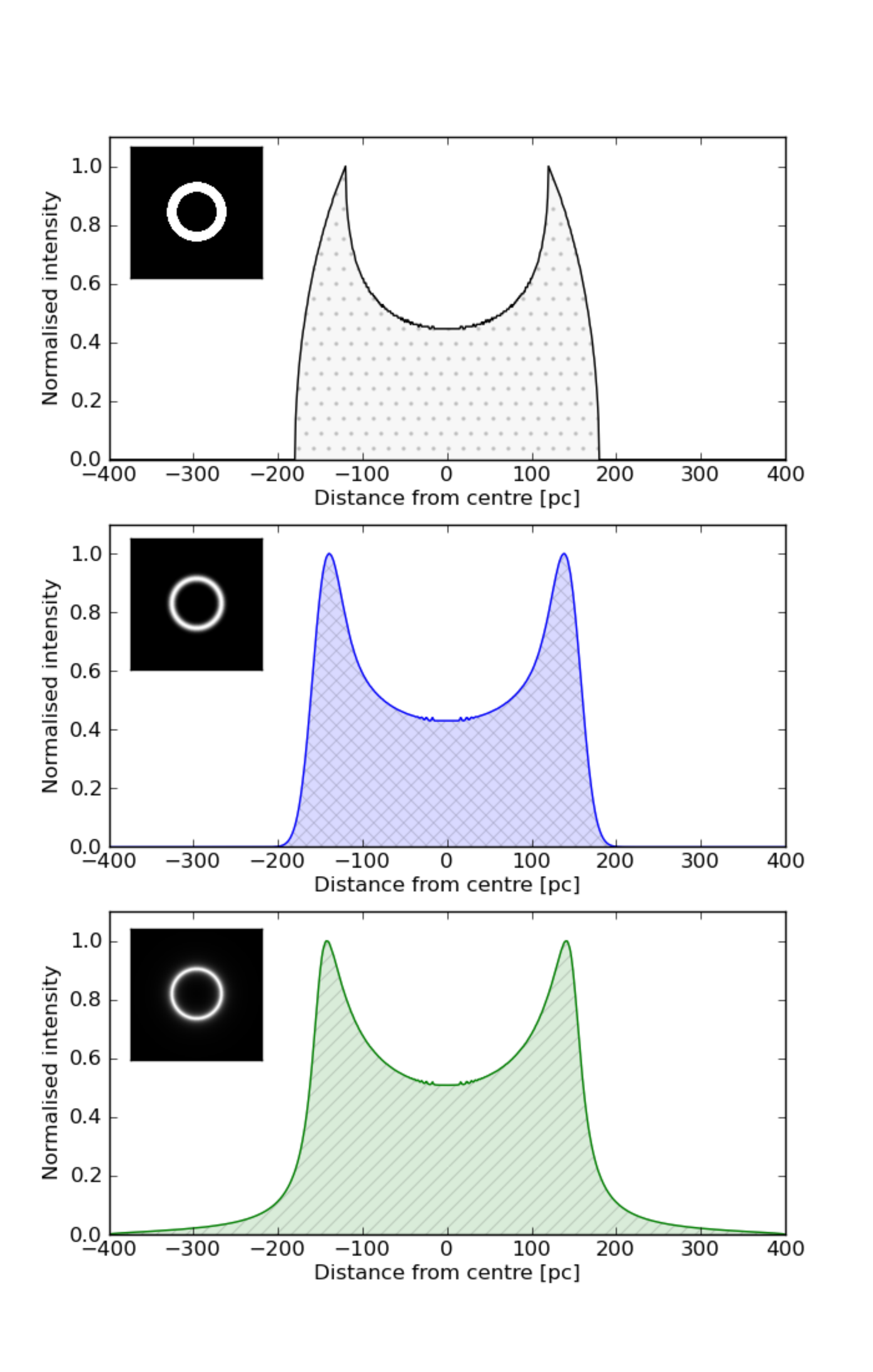}
\caption{Model of clear shell number 1. The profiles obtained by a slit placed in front of the centre of the shell. {\it Upper panel:} Constant density. {\it Middle panel:} Gaussian density. {\it Bottom panel:} Cauchy-Lorentz density. The stamps are 800\,pc wide.} \label{fig:modelShell}
\end{figure}

In order to test this hypothesis, we performed simple geometrical models, beginning with the simplest assumption: an empty sphere bounded by a spherical shell of constant width and density. The profile of constant density is shown in Fig.~\ref{fig:profileDensity}, left profile, in grey. The centre of the envelope is set at a radius of 150\,pc from the centre of the shell and its full width is 60\,pc. In the optically thin case, the profile obtained by a slit placed in front the centre of the shell is shown in the upper panel of Fig.~\ref{fig:modelShell}, where the profile has been normalised to its maximum intensity. The first striking result is how well this very simple model reproduces the main features observed in the data (see the upper panel of Fig.~\ref{prof_clearshell1}). In particular, the central dip feature, a decrease of roughly 50\% with respect to the peaks, seen in the model is also seen in all the shells and clear shells in our catalogues. This is a direct consequence of the spherical assumption for the geometry of the shell and envelope, and it is the first time that this has been proved on such an amount of shells and clear shells.

Nevertheless, this first result shows that the assumption of a constant density profile is only an approximation. Indeed, the two horns of the profile show very steep departures at the locations of the outer boundaries of the envelope, as well as very sharp turnarounds at the maxima, i.e. the location of the inner boundary of the envelope. Both features are due to the discrete form of the function used for the constant density profile. In order reproduce more physical shells, these very sharp edges for the envelope need to be smoothed. For instance, a density profile following a Gaussian (normal) distribution could be more appropriate and realistic. This Gaussian function reads:
\begin{equation} \label{eq:gauss}
\qquad \qquad \qquad f(x) = \frac{1}{\sqrt{2\pi\sigma^2}} e^{ -\frac{(x-\mu)^2}{2\sigma^2} }
\end{equation}
where parameter $\mu$ is the mean (location of the peak, at a radius of 150\,pc from the centre of the shell) and $\sigma$, 15\,pc, is the standard deviation (i.e. a measure of the full-width at half-maximum of the distribution is $\mathrm{FWHM} = 2\sqrt{2 \ln 2 }\;\sigma \approx 35\,$pc). The central (blue) profile in Fig.~\ref{fig:profileDensity} shows its density distribution. The profile obtained by a slit placed in front the centre of the shell is shown in the middle panel of Fig.~\ref{fig:modelShell}. Qualitatively, the observed profile is similar to the previous one but the profile looks more like the real one, showing a smoother distribution. However, in the profile obtained from the data, the outer boundaries of the envelope still appear less sharp, displaying two very well marked wings which are not well reproduced by the Gaussian density distribution.

In order to reproduce this feature, while keeping the smooth profile obtained with the Gaussian density profile, we can use a Cauchy-Lorentz distribution, which reads:
\begin{equation} \label{eq:cauchy}
\qquad \qquad \qquad f(x) = { 1 \over \pi } \left[ { \gamma \over (x - x_0)^2 + \gamma^2 } \right]
\end{equation}
where $x_0$ is the location parameter, specifying the location of the peak of the distribution (a radius of 150\,pc from the centre of the shell), and $\gamma$ is the scale parameter which specifies the half-width at half-maximum (i.e. 15\,pc) in Eq.~\ref{eq:cauchy}. Following the Cauchy-Lorentz profile, this translates into a full width of 60\,pc at a level of 20\% of its maximum, where 80\% of the flux is enclosed. The larger extension of the Lorentz distribution succeeds in reproducing the wings seen in the integrated profile (see the bottom panel of Fig.~\ref{fig:modelShell}). A side effect of the large extension of the wings is that more matter accumulate also towards the centre of the shell and, as a consequence, the centre of the profile reaches a level of 50\% of the peaks, while it was 42\% for the Gaussian distribution. 

\begin{figure}
\includegraphics[width=\columnwidth]{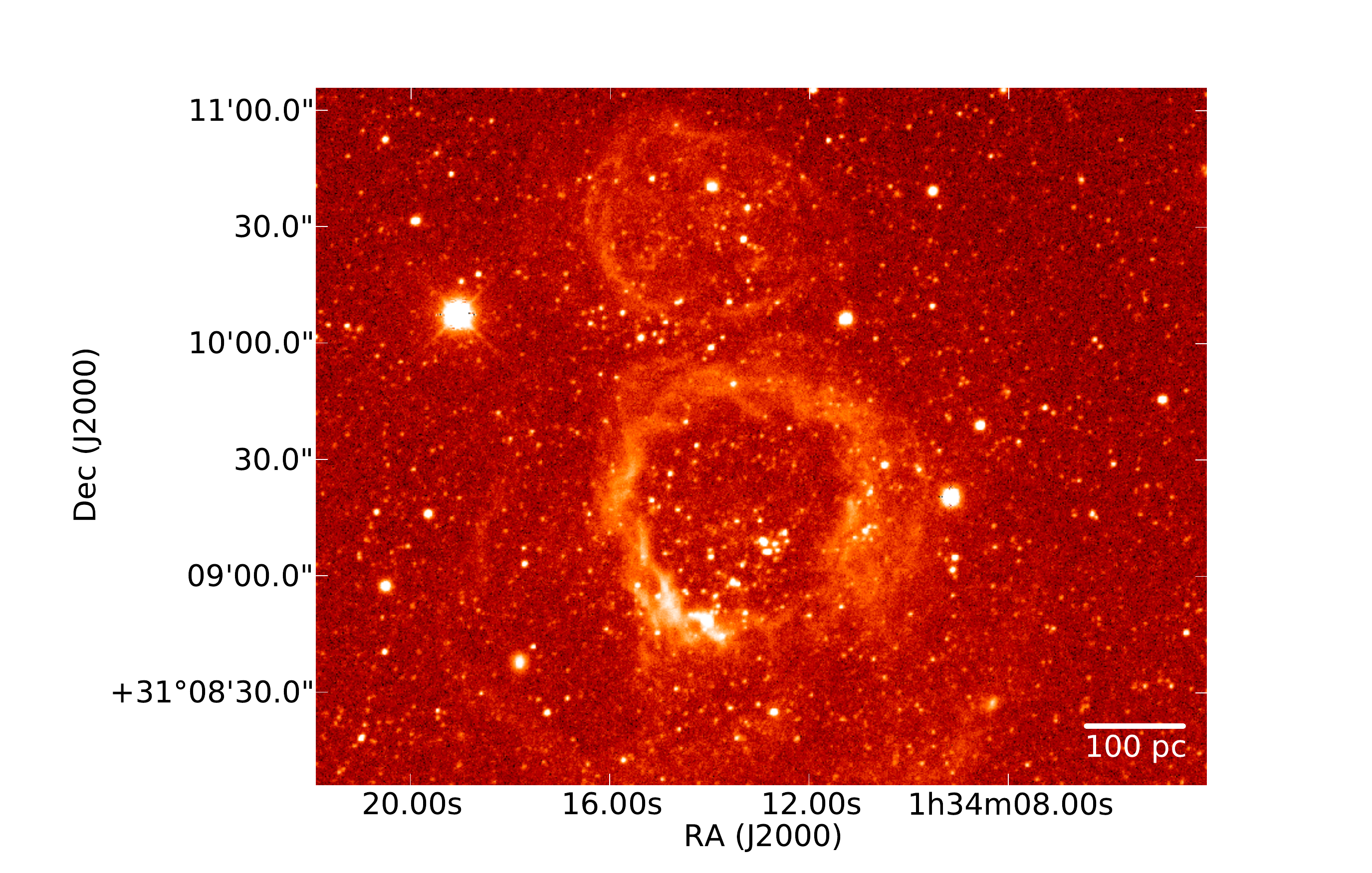}
\caption{H$\alpha$ image of clear shell number 87 \citep{Massey:2006p517}.} \label{fig:clearShell1Ha}
\end{figure}

We have to note that the images have been degraded to the {\it Herschel} SPIRE 250\,$\mu$m resolution, so the observed profiles are only indicating of the general behaviour of the density distribution. We verified that the degradation to a lower resolution did not change significantly the loci of the shell boundaries and maxima. In fact, examining the best resolution images currently available in H$\alpha$ \citep[see Fig.~\ref{fig:clearShell1Ha}, Local Group Survey, from][]{Massey:2006p517} shows that the envelope is composed of lots of filaments superimposed on the underlying distribution that we have modelled. In almost all the other shells, while looking at high resolution, we can distinguish various envelopes with different radii, and generally centred on the same UV stellar clusters. The ubiquity of these features could indicate that clusters of new stars trigger periodically new generations of star formation in envelopes around them, but more detailed models are needed to corroborate this hypothesis.

\subsection{Electron density of clear shells and filled regions} \label{subsec:profiles_ne}
As we have shown in the previous section that the geometry of the clear shells can be well represented by a spherical shell, we can exploit this feature in order to infer more physical conditions about the shells. For instance, as we know the length of the emission distribution, we can infer the electron density in the envelope. To do so, we selected only the five clear shells which had a good enough homogeneous and projected circular shape in order to be able to reasonably estimate the size of the emission distribution along the line of sight. The emission measure (EM) has been measured in the very centre of the clear shell, and the size of the emitting regions has been estimated by adding the two thickness of the clear shell in the outer west and east parts, and we obtain
the electron density, $n_e$, using a constant and a Cauchy-Lorentz profile. We list in Table~\ref{tab:neShells} the constant electron density, $n_{e\ {\rm const.}}$, as well as the maximum of the Cauchy-Lorentz distribution of the electron density, $n_{e\ {\rm max.}}$, for five clear shells.

\begin{table}
\caption{Electron density of clear shells. The constant electron density, $n_{e\ {\rm const.}}$, as well as the maximum of the Cauchy-Lorentz distribution of the electron density, $n_{e\ {\rm max.}}$, are shown in columns 4 and 5, respectively. Electron density upper-limits in the last column ($n_{e\ {\rm spec.}}$) are from spectroscopic data \citep{Magrini:2007p670}.}
\label{tab:neShells}
\centering
\begin{tabular}{l c c c c c}
\hline\hline
Source  & L & Emission Measure & $n_{e\ {\rm const.}}$ & $n_{e\ {\rm max.}}$ & $n_{e\ {\rm spec.}}$ \\
 Number & [pc] & [pc\,cm$^{-6}$] & [cm$^{-3}$] & [cm$^{-3}$] & [cm$^{-3}$] \\
\hline
87    & 120 & 13 & 0.3 & 0.4 & - \\
48    & 110 & 40 & 0.6 &  0.8 & - \\
32  & 105 & 26 & 0.5 &  0.7 & $<30$ \\
66  & 100 & 50 & 0.7 &  1.0 & - \\
116   & 85  & 14 & 0.4 &  0.6 & - \\

\hline
\end{tabular}
\end{table}

Unlike for the shell study, with respect to the filled regions, we do not know the size of emission distribution along the line of sight.
As it is not possible to directly measure the linear size of the line of emission, we concentrated on rather filled, circular regions and add the hypothesis that the \hii\ region is spherical in three dimensions. Hence the full extent from west to east will provide us an estimation of the linear size of the emission along the line of sight. This hypothesis has already been widely used in previous studies \citep{Magrini:2007p670,Esteban:2009p563}. In Table~\ref{tab:neFilled}, we list the obtained electron density of 11 filled regions.

\begin{table}
\caption{Electron density of filled regions. Electron density upper-limits in the last column ($n_{e\ {\rm spec.}}$) are from spectroscopic data \citep{Magrini:2007p670}.}
\label{tab:neFilled}
\centering
\begin{tabular}{l c c c c}
\hline\hline
Source  & L & Emission Measure & $n_e$ & $n_{e\ {\rm spec.}}$ \\
Number & [pc] & [pc\,cm$^{-6}$] & [cm$^{-3}$] & [cm$^{-3}$] \\
\hline
23& 175 & 970  & 2.4 & $<20$\\
35& 220 & 83   & 0.6 & - \\
59& 160 & 42   & 0.5 & - \\
77& 180 & 222  & 1.1 & $<10$ \\
47& 180 & 411  & 1.5 & - \\
102& 175 & 156  & 0.9 & $<10$ \\
57& 170 & 222  & 1.1 & - \\
34& 185 & 100  & 0.7 & - \\
5& 190 & 211  & 1.1 & $<10$ \\

\hline
\end{tabular}
\end{table}

The electron densities for filled and clear shell \hii\ regions are comparable. Nevertheless, the electron density in filled \hii\ regions may be two to five times higher with respect to the envelope of the shells. This may be expected in the sense that filled regions are more compact that the shells for which the large space covered may have decreased the electron density.

We compared our results with works which have been carried in the literature for the same galaxy, employing different methods (using mainly spectroscopic data) and obtained consistent results. \citet{Magrini:2007p670} obtained spectroscopically the electron density for 72 emission-line objects, including mainly \hii\ regions. We have 5 \hii\ regions in common with their work. Although they could only give upper limits for the majority of their regions, we find that our results are statistically compatible (they found that the majority of their \hii\ regions have $n_e < 10$), and in particular we are able to give the electron densities for 5 regions, in agreement with the upper limits given by \citet{Magrini:2007p670} for these 5 sources. The five \hii\ regions in common are the filled regions number 23, 77, 102, 5 and the clear shell number 32, which correspond to their sources CPSDP~194, VGHC~2-84, BCLMP~717b, BCLMP~238, and M33SNR~25, respectively. The upper limits they have obtained are listed in Tables~\ref{tab:neShells} and \ref{tab:neFilled} for the corresponding sources. 

\citet{Esteban:2009p563} derived the electron density for the bright \hii\ regions NGC\,595 and NGC\,604 from Keck~I spectrophotometric data. For NGC\,595, they found electron densities of $270\pm180$ from [N\,{\sc i}], $260\pm30$ from [O\,{\sc ii}], $700\pm370$ from [Cl\,{\sc iii}], and minor than 100 from [S\,{\sc ii}]. For NGC\,604, they found electron densities of $140\pm100$ from [N\,{\sc i}], $270\pm30$ from [O\,{\sc ii}], $490^{+510}_{-490}$ from [Cl\,{\sc iii}], and minor than 100 from [S\,{\sc ii}].

We note that the spectroscopic values for the electron densities in the works by \citet{Magrini:2007p670} and \citet{Esteban:2009p563} are not directly comparable with the electron densities obtained from our photometric data. The electron density derived from spectroscopic observations corresponds to the density of the clumps within the region. Here, we derive a mean value for the electron density (r.m.s. density) for the whole envelope. Moreover, in a photometric sample of \hii\ regions in the galaxy NGC\,1530, \citet{Relano:2005p644} found mean electron densities for the shells of typically 10\,cm$^{-3}$, one order of magnitude higher than the ones obtained in M\,33. The difference comes from the size considered for the envelope: while we consider the full extension of the envelope, \citet{Relano:2005p644} take only into account the size of the brightest filaments, 4.5\,pc. The compact and clear shell regions follow the size-density relation for extragalactic \hii\ regions described by \citet{2009A&A...507.1327H} and \citet{2011ApJ...732..100D} for dust \hii\ regions. With a diameter of the order of 100\,pc and densities of the order of 1\,cm
$^{-3}$, the \hii\ regions considered in the present study are among the largest and least dense within the dynamical range displayed in the
Fig.~2 of \citet{2009A&A...507.1327H} and in Fig.~11 of \citet{2011ApJ...732..100D}. Indeed, they match more particularly the zone defined by the \hii\ regions of nearby galaxies
studied by \citet{1984ApJ...287..116K}.

\section{Discussion}\label{sec:disc}

\hii\ regions are formed of hydrogen that has been ionised by the radiation coming from the central massive stars. During their short life, massive stars are able to generate the radiation that ionise the hydrogen \citep{Vacca:1996p794,Martins:2002p795} but they also emit stellar winds that interact with the gas surrounding them \citep{Dyson:1979p797,Dyson:1980p798,Kudritzki:2002p796}. The effect of the stellar winds in the region is to create a cavity or shell structure, sweeping the gas around the stars with typical expansion velocities of around $\sim$50\,km\,s\me\ \citep{Chu:1994p791,Relano:2005p644}. There is also evidence that the ionised shells are related to shell structures in the IR bands, which trace the emission of the dust \citep[e.g.][]{Watson:2008p577,Verley:2010p687} and in some cases the compression of the interstellar gas surrounding the shell can produce new star formation events called triggered star formation \citep{1987ApJ...317..190M,2001AJ....122.3017S}. Observational evidence of triggered star formation based on data from {{\it Herschel}} has been recently found in Galactic \hii\ regions \citep[e.g.][]{2010A&A...518L.101Z,2010A&A...518L..81Z}.

The mass-loss rate of the massive stars increases at $\sim$3\,Myr and is maintained at a high rate till $\sim$40\,Myr, when it decays very strongly \citep{Leitherer:1999p491}. During this time supernova explosions can occur adding more kinetic energy to the previously formed shell. In this scenario, in a young and already formed \hii\ region the central stars will be ionising the gas around them but their stellar winds have not had time to produce a shell of swept gas. In this case the \hii\ region would look like a filled knot of ionised gas emitting at \ha. After $\sim$3-4\,Myr the bubble will be created by the stellar winds and a shell of ionised gas will expand into the ISM. The size and the expansion velocity of the shell will depend on the amount of kinetic energy provided by the central stars and the evolution will depend on the physical conditions of the ISM surrounding it. If the gas in the ISM is tenuous and the pressure is low, then the shell will be able to expand more easily and large shell structures could be created. \citet{Whitmore:2011p726} has recently suggested a relation between the region morphology and the age of the central cluster: very young (less than a few\,Myr) clusters would show the \ha\ emission of the ionised gas coincident with the cluster stars, clusters $\approx$\,5\,Myr would have the gas emission located in small shell structures around the stars, and in still older clusters ($\approx$\,5-10\,Myr ) the \ha\ emission would show even larger shell structures. If no \ha\ emission is associated with the cluster this should be older than $\approx$\,10\,Myr.

We can give an estimate of the age of the clear shells based on the kinematics of the ionised gas. Using \ha\ Fabry-Perot spectroscopy \citet{Relano:2005p644} found that the \ha\ emission line profiles of the \hii\ region populations of three late-type spiral galaxies show evidence of a shell of ionised gas expanding in the ISM with expansion velocities range from 40 to 90\,km\,s\me\ and mean values of $\sim$50-60\,km\,s\me. Assuming an expansion velocity of $\sim$50\,km\,s\me\ for the ionised shells and taking into account a typical radius for the observed shells of $\sim$200\,pc (see Fig.~\ref{histo}), we derive a kinematic age for the cluster of $\sim$4\,Myr, which agrees with the timescale provided by \citet{Whitmore:2011p726}. 

We also find evidence of a secondary generation of stars within the shells. In the top-left panel of Fig.~\ref{prof_clearshell1}, we see the emission distribution of \ha\ and FUV for one example of a clear shell: \ha\ tracing the boundary of the shell with the two horns of emission and a centred peak of FUV (and also NUV) corresponding to the stellar clusters. However, we also see in the right peak of \ha\ emission a knot of FUV (and NUV), which would correspond to the emission of new born stars within the shell. This \hii\ region might be old enough to have triggered star formation within the swept shell and therefore part of the shell is being ionised by this secondary generation of stars. This phenomenology is observed in 12 clear shells. Some cases are very clear while others (7 out of 12) are marginal detections. The marginal detections depend on the geometry of the clear shell, as not all the clear shells are completely spherical, and the exact location where the profiles have been extracted.

\section{Summary and conclusions}\label{sec:sum}

We select 119 \hii\ regions in the local group spiral galaxy M\,33, and classify them according to their morphology: filled, mixed, shell, and clear shell \hii\ regions. Using a multi-wavelength set of data, from FUV (1516\,\AA) to IR (SPIRE 250\,\mi), we study their SED, the influence of the stellar radiation field, and their intensity profiles. Here are our main conclusions:

\begin{itemize}

\item An analysis of the SED of each region shows that regions belonging to one group show roughly the same features. Besides, these features are different from one classified group to another, showing that the dominant physical processes vary among the different morphologies. The SED,  normalised to the emission at 24\,\mi, shows that FIR peak for shells and clear shells seems to be located towards longer wavelengths, indicating that the dust is colder for this type of object. In the SED normalised to the FUV flux filled regions present the highest FIR fluxes, which shows that in these regions the dust is so close to the central stars that it is very efficiently heated, while shells and clear shells present less flux in this normalisation because the dust is in general distributed further away from the central stars in these regions.

\item The warm dust colour temperature traced by the 100\,\mi/70\,\mi\ ratio shows that two regimes are in place. The filled and mixed regions show a well constrained value of the logarithmic 100\,\mi/70\,\mi\ ratio over one order of magnitude in \ha\ and FUV surface brightness, while the shells and clear shells show a wider range of values of this ratio of almost two orders of magnitude. The spatial relation between the stars and dust could have an effect in the heating mechanism of the dust in the \hii\ regions.
 
\item We estimate the dust mass fitting \citetalias{Draine:2007p588} models to the SED of each individual region. We find dust masses within the range $10^{2}-10^{4}\msun$, consistent with those derived for \hii\ regions in other galaxies using the same models. The 250\,\mi/160\,\mi\ ratio, an estimator of the temperature of the cold dust, shows that shells and clear shells tend to have lower cold dust temperatures than mixed and filled regions. An estimate of the cold dust temperature for each \hii\ region using the same dust models gives a range of $\rm T_{cold}\sim12-27\,K$ for the whole sample.

\item For all the wavelength bands available in this study we extract East-West and North-South profiles of each \hii\ region individually. For filled regions the emission at all bands occurs at the same location. On the contrary, clear shell regions generally show the two horns describing the shells. For some cases, we find evidence that current star formation takes place within the envelope (possibly due to triggered star formation processes).

\item Concentrating on the \ha\ emission, we propose that the clear shells that appear circular on the plane of the sky are the result of the line-of-sight projection of three-dimensional spherical shells. We find that the density within the envelope of the shells follows more closely a Cauchy-Lorentz distribution rather than constant or Gaussian density distributions. Nevertheless, high resolution images show that this averaged density distribution is in fact the result of lots of high and low density filamentary structures.

\item Knowing the real \ha\ structure in three dimensions allows us to measure the electron density in the clearest shells, because we are able to estimate the length of the emission distribution. From five clear shells, the mean electron density is $n_e = 0.7\pm0.3\,{\rm cm^{-3}}$. The electron density in the clear shells is hence rather comparable to the one we obtained from eleven filled regions, although some filled regions could reach electron densities 2 to 5 times higher than the one found in the envelope of clear shells. As we are using photometric data, these electron density estimations are only global averages because the bright filaments within the envelope would result in higher electron densities with respect to the faintest zones in between. Nevertheless, our estimations are compatible with spectroscopic work done in the literature \citep{Magrini:2007p670,Esteban:2009p563} for \hii\ regions in M\,33.

\end{itemize}

\begin{acknowledgements}

Part of this research has been supported by the ERG HER-SFR from the EC.  This work was partially supported by a Junta de Andaluc\'ia Grant FQM108, a Spanish MEC Grant AYA-2007-67625-C02-02, and Juan de la Cierva fellowship Program. This research made use of APLpy, an open-source plotting package for
Python hosted at http://aplpy.github.com ; of TOPCAT \& STIL: Starlink
Table/VOTable Processing Software \citep{2005ASPC..347...29T}; of
Matplotlib \citep{Hunter:2007}, a suite of open-source python modules
that provides a framework for creating scientific plots.

 \end{acknowledgements}

\bibliographystyle{aa} 
\bibliography{mnras}  

\clearpage
\appendix
\section{Two-dimensions multi-wavelength projected profiles of \hii\ regions}
\begin{figure*}
\centering
\includegraphics[width=\columnwidth]{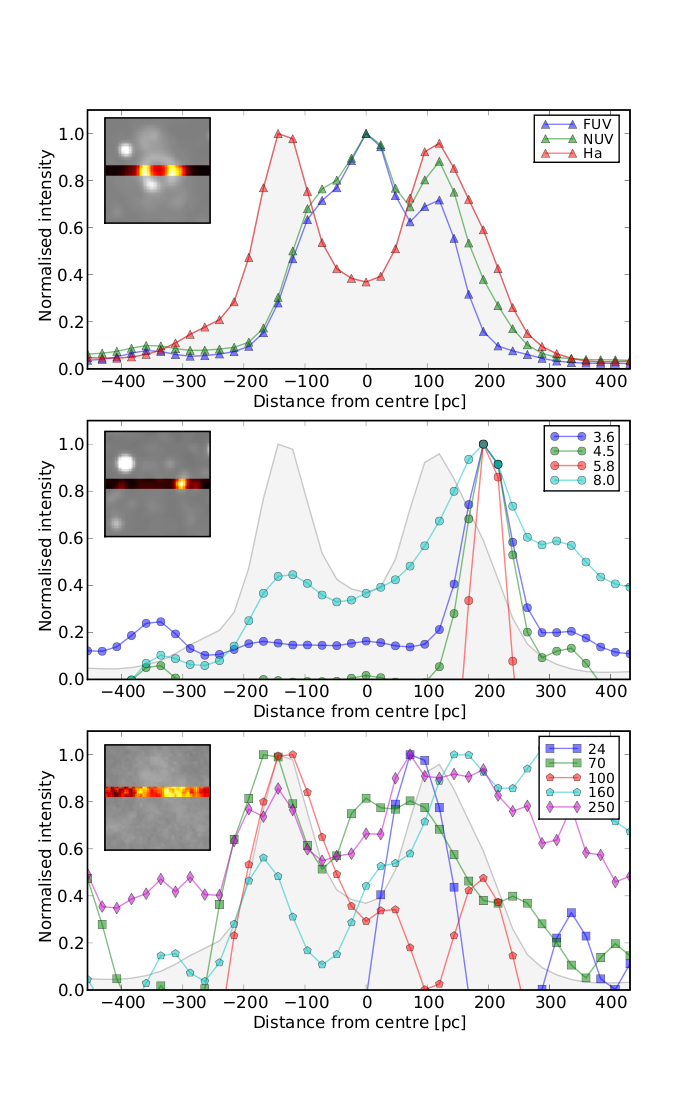}   
\includegraphics[width=\columnwidth]{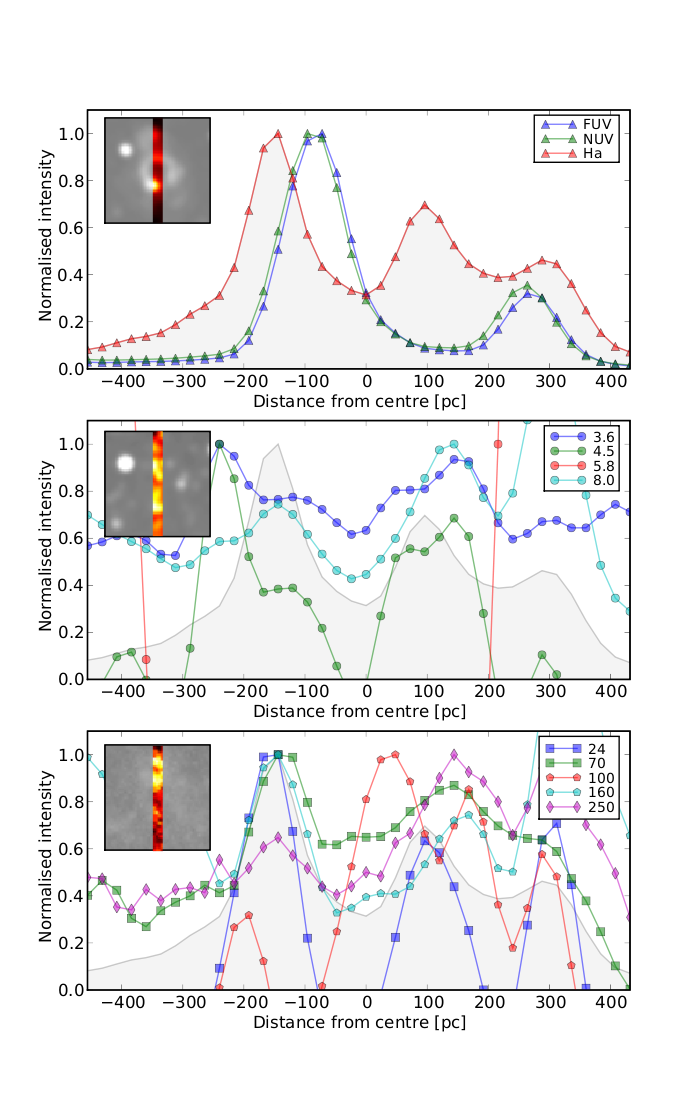}   
\caption{Emission line profiles for \hii\ region~87 classified as clear shell in the horizontal (left) and vertical (right) direction. For clarity, the profiles are separated into three panels ({\it top:} \ha, FUV, NUV; {\it middle:} 3.6\,\mi, 4.5\,\mi, 5.8\,\mi, 8.0\,\mi; {\it bottom:} 24\,\mi, 70\,\mi, 100\,\mi, 160\,\mi, 250\,\mi). The small box at top-left corner in each panel shows the location of the profile overlaid on the region, the images correspond to \ha, 4.5\,\mi, and 250\,\mi\ for the top, middle, and bottom panels, respectively. All the profiles are normalised to their maxima. The \ha\ profile is depicted in grey in all the panels for reference.}
\label{prof_clearshell1}
\end{figure*}

\begin{figure*}
\centering
\includegraphics[width=\columnwidth]{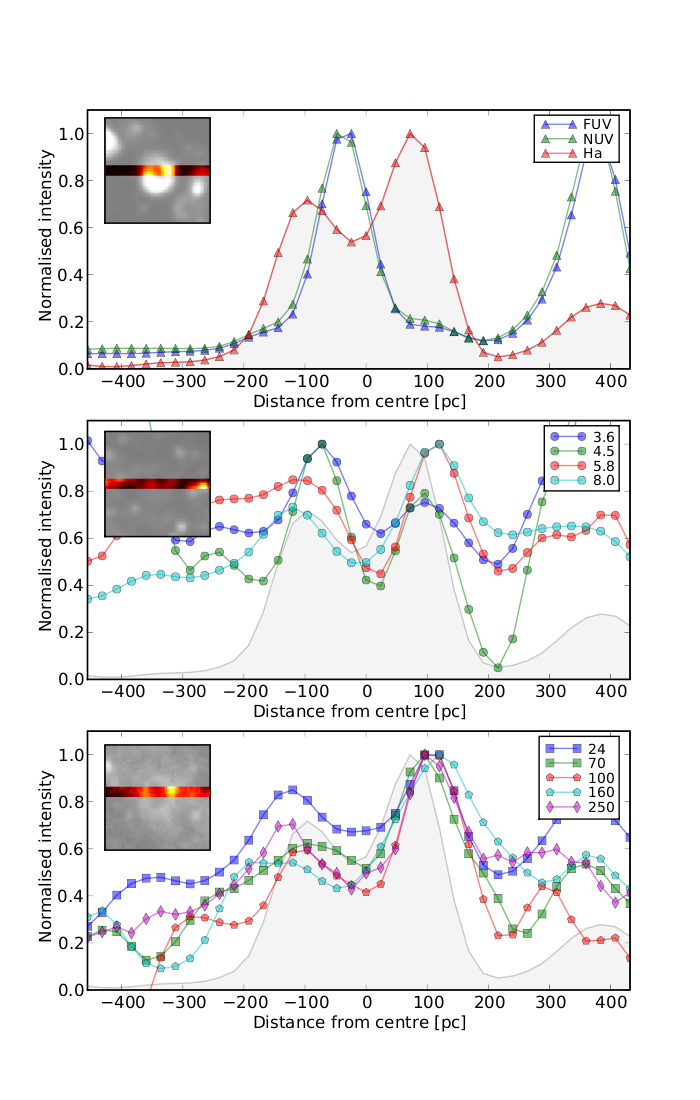}   
\includegraphics[width=\columnwidth]{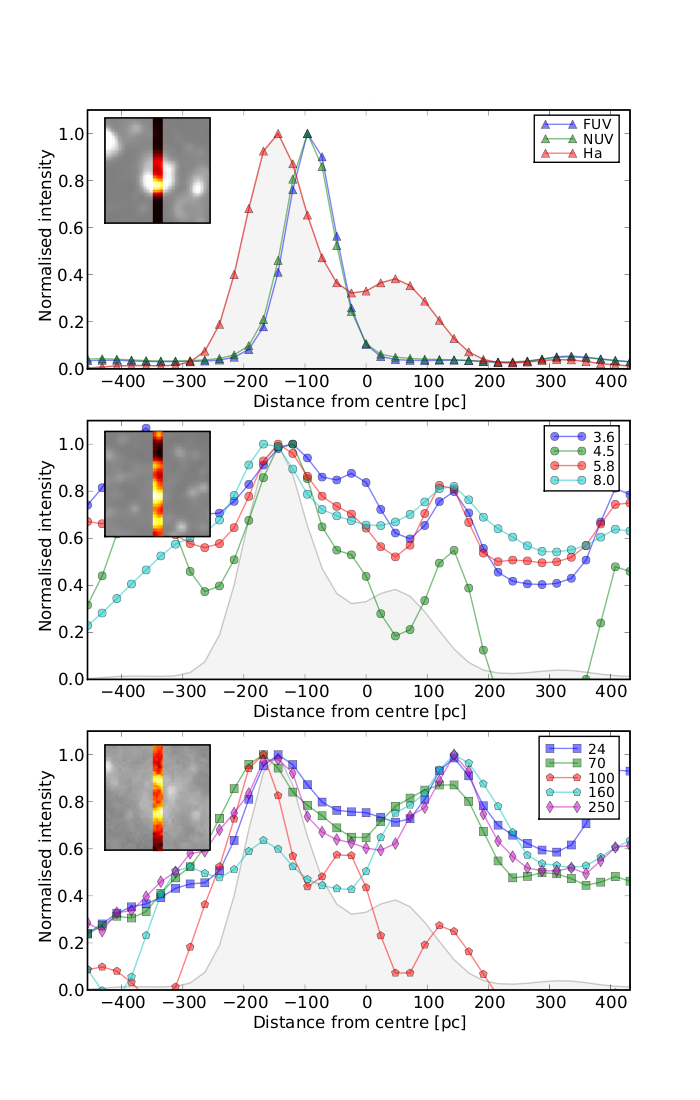}   
\caption{Same as Fig.~\ref{prof_clearshell1} for \hii\ region number~48 classified as clear shell.}
\label{prof_clearshell3}
\end{figure*}

\begin{figure*}
\centering
\includegraphics[width=\columnwidth]{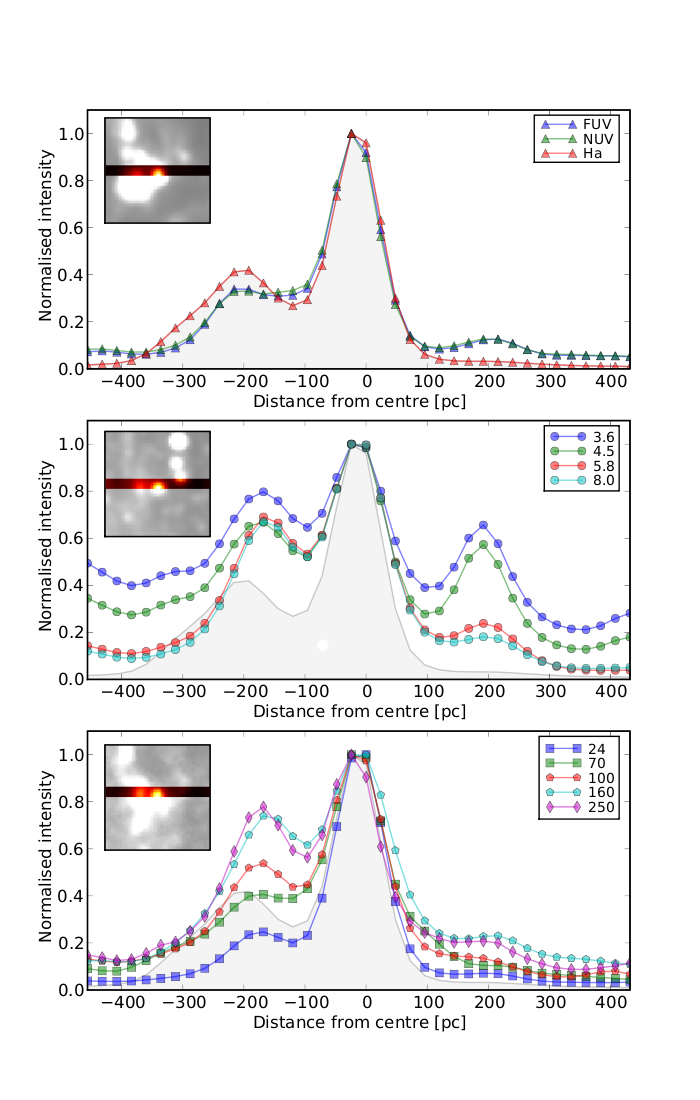}   
\includegraphics[width=\columnwidth]{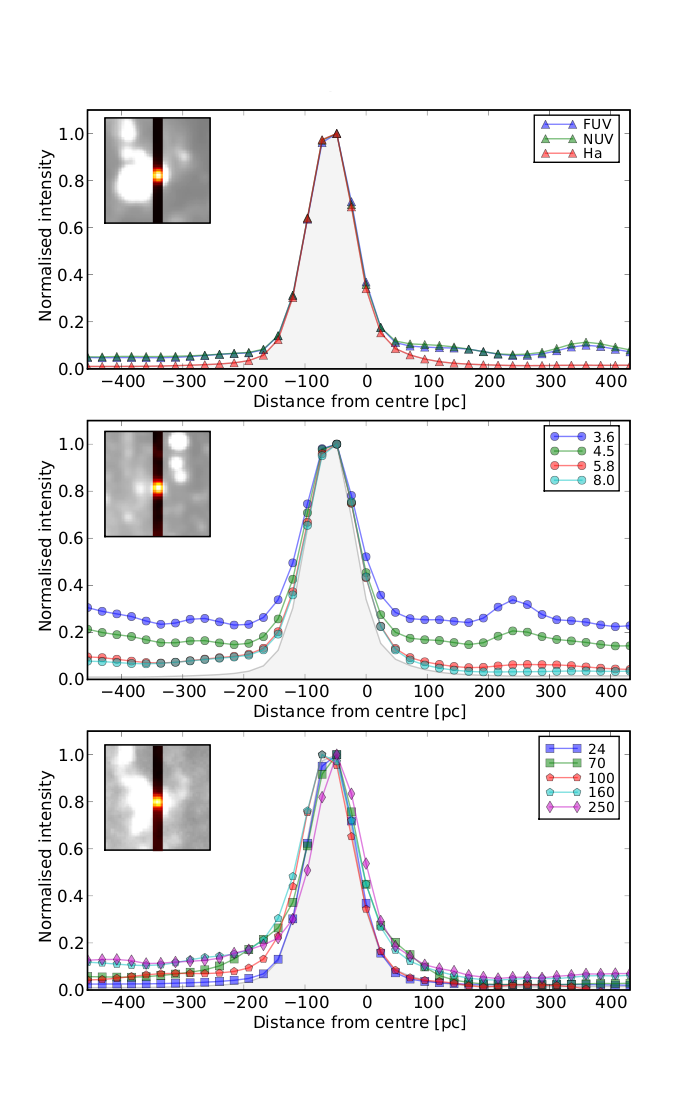}   
\caption{Same as Fig.~\ref{prof_clearshell1} for \hii\ region number~23 classified as filled.}
\label{prof_compact20}
\end{figure*}

\begin{figure*}
\centering
\includegraphics[width=\columnwidth]{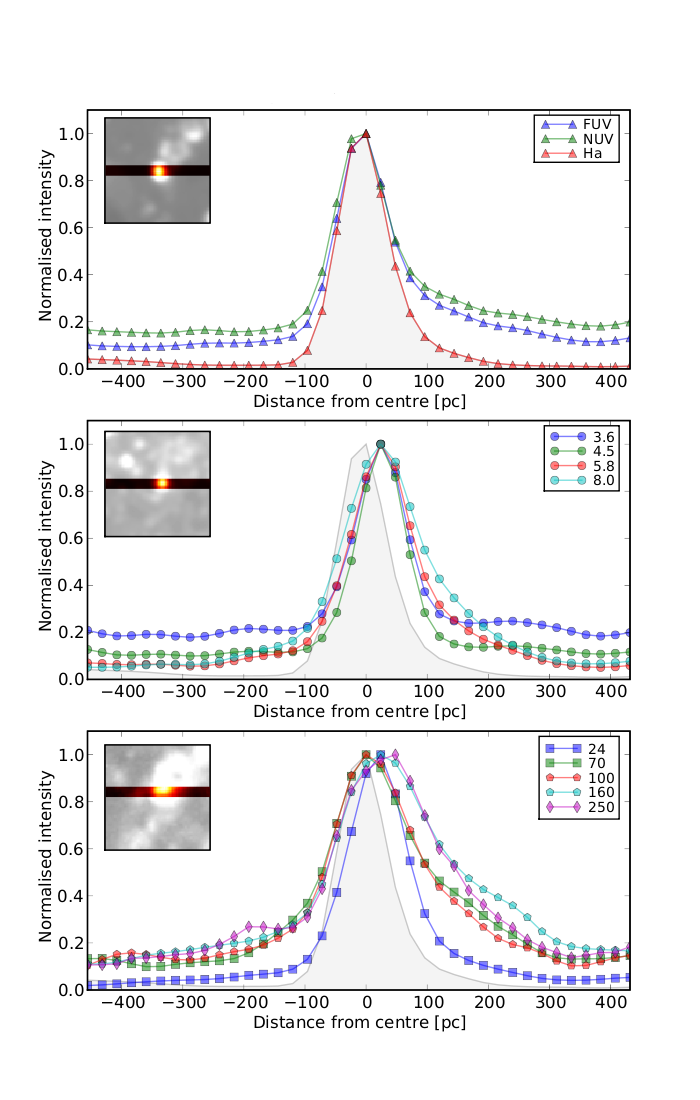}   
\includegraphics[width=\columnwidth]{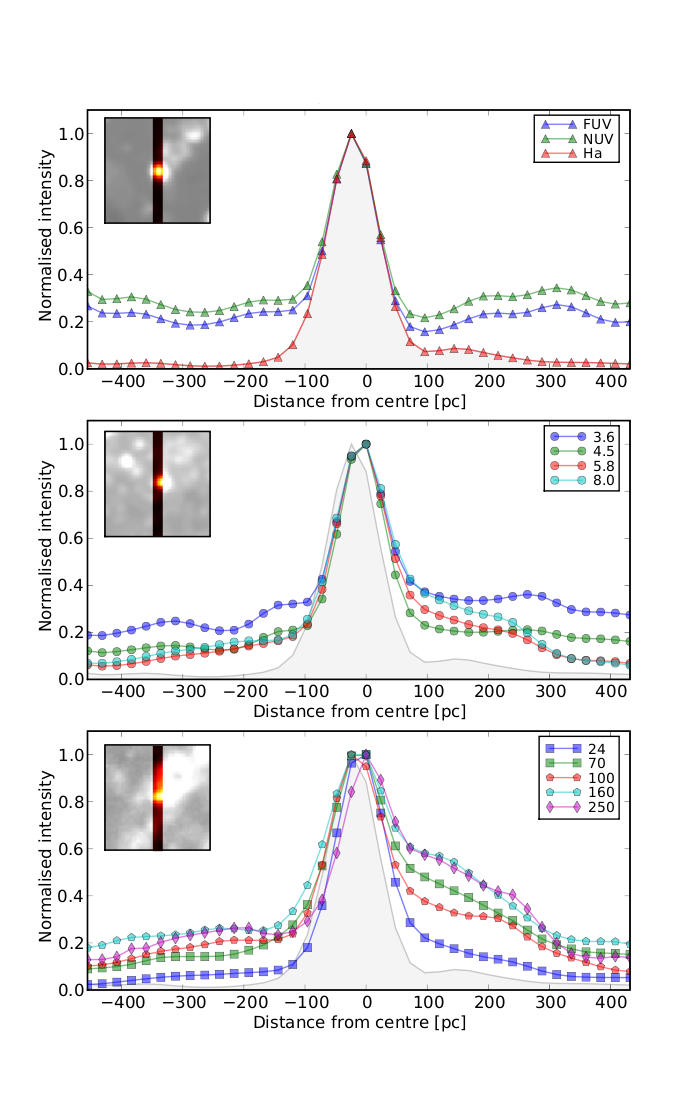}   
\caption{Same as Fig.~\ref{prof_clearshell1} for \hii\ region number~35 classified as filled.}
\label{prof_compact39}
\end{figure*}

\section{Tables}

\end{document}